\documentclass[iop]{emulateapj}
\usepackage{amsmath}
\usepackage{amssymb}
\usepackage{mathbbol}
\usepackage{dsfont}

\DeclareGraphicsExtensions{.eps,.eps.gz,.epsi}

\newcommand{\teff}{\mbox{$T_{\rm eff}$}} \newcommand{\logg}{{\rm{log}~$g$}}
\newcommand{\feh}{{\rm [Fe/H]}}

\shorttitle{Stellar loci II. a model-free estimate of the binary fraction for field FGK stars}
\shortauthors{Yuan et al.}

\begin{document}

\title{Stellar loci II. a model-free estimate of the binary fraction for field FGK stars}

\author
{Haibo Yuan\altaffilmark{1, 2},
Xiaowei Liu\altaffilmark{3, 1},
Maosheng Xiang\altaffilmark{3},
Yang Huang\altaffilmark{3},
Bingqiu Chen\altaffilmark{3},
Yue Wu\altaffilmark{4},
Yonghui Hou\altaffilmark{5},
Yong Zhang\altaffilmark{5}
}
\altaffiltext{1}{Kavli Institute for Astronomy and Astrophysics, Peking University, Beijing 100871, P. R. China; email: yuanhb4861@pku.edu.cn}
\altaffiltext{2}{LAMOST Fellow}
\altaffiltext{3}{Department of Astronomy, Peking University, Beijing 100871, P. R. China; email: x.liu@pku.edu.cn}
\altaffiltext{4}{Key Laboratory of Optical Astronomy, National Astronomical Observatories, Chinese Academy of Sciences, Beijing 100012, P. R. China}
\altaffiltext{5}{Nanjing Institute of Astronomical Optics \& Technology, National Astronomical Observatories, Chinese Academy of Sciences, Nanjing 210042, P. R. China}

\journalinfo{submitted to The Astrophysical Journal}
\submitted{Received ; accepted }
 
\begin{abstract}
We propose a Stellar Locus OuTlier (SLOT) method to determine the binary
fraction of main-sequence stars statistically.  The method is sensitive to
neither the period nor mass-ratio distributions of binaries, and able to
provide model-free estimates of binary fraction for large numbers of stars of
different populations in large survey volumes. We have applied the SLOT method
to two samples of stars from the SDSS Stripe\,82,  constructed by combining the
re-calibrated SDSS photometric data with respectively the spectroscopic
information from the SDSS and LAMOST surveys. For the SDSS spectroscopic
sample, we find an average  binary fraction for field FGK stars of 41\%$\pm$2\%.
The fractions decrease toward late spectral types, and are respectively
44\%$\pm$5\%, 43\%$\pm$3\%, $35\%\pm5\%$, and $28\%\pm6\%$ for stars of $g-i$ colors between
0.3 -- 0.6, 0.6 -- 0.9, 0.9 -- 1.2, and 1.2 -- 1.6\,mag.  A modest metallicity
dependence is also found. The fraction decreases with increasing metallicity.
For stars of \feh between $-0.5$ -- 0.0, $-1.0$ -- $-0.5$, $-1.5$ -- $-1.0$,
and $-2.0$ -- $-1.5$\,dex, the inferred binary fractions are $37\%\pm3\%$,
$39\%\pm3\%$, $50\%\pm9\%$, and $53\%\pm20\%$, respectively. We have further divided the
sample into stars from the thin disk, the thick disk,  the transition zone
between them, and the halo.  The results suggest that the Galactic thin and
thick disks have comparable binary fractions, whereas the Galactic halo
contains a significantly larger fraction of binaries. Applying the method to
the LAMOST spectroscopic sample yields consistent results. Finally, other
potential applications and future work with the method are discussed.
\end{abstract}

\section{Introduction} 

Binaries are ubiquitous in the universe (e.g., Heintz 1969; Abt \& Levy 1976; Duquennoy \& Mayor 1991; Raghavan et al. 2010).
They play an important role in our understanding of the star formation and evolution,
such as the production and properties of the blue stragglers, cataclysmic variables, X-ray binaries,
type Ia supernovae, and planetary nebulae. 
They also play an active role in the studies of stellar population synthesis (e.g., Hurley et al. 2002; Li \& Han 2008). 
However, it is very challenging to discriminate a binary system from a single star
and to characterize the binary stars (e.g., the binary fraction\footnote{In the current work, the binary fraction 
is defined as the fraction of stars that have a secondary companion.}, 
denoted as $f_{b}$, and the distributions of mass ratios, orbital periods, and eccentricities) 
for different stellar populations and Galactic environments (see the review by Duch{\^e}ne \& Kraus 2013 and references therein).

A variety of techniques, visual, spectroscopic, photometric, and astrometric, 
have been developed  to search for binaries, leading to respectively the discoveries of 
visual, spectroscopic, eclipse, and astrometric binaries.
Spectroscopic, eclipse, and astrometric binaries are found by periodic changes in the stellar radial velocity,
brightness, and positions, respectively.
All those techniques are biased in one way or other, and in many cases limited to bright stars in the solar neighborhood. 
For example, spectroscopic and photometric techniques are biased to close
binaries consisting of stars of similar masses. Both techniques are also time consuming. 
Visual binaries discovered by high angular resolution imaging techniques such as speckle interferometry 
are biased to long-period, similar-mass binaries.
Astrometric binaries are biased to binaries of intermediate periods and currently limited to bright stars observed with the Hipparcos.
With the advent of millions of stellar spectra, collected by for example, the Sloan Digital Sky Survey (SDSS; York et al. 2000)
and the LAMOST (Cui et al. 2012) spectroscopic surveys (Zhao et al. 2012; Deng et al. 2012; Liu et al. 2014), spectral binaries 
have also been identified as those whose spectra show two distinct sets of photospheric features, such as the 
white-dwarf-main-sequence binaries (e.g., Rebassa-Mansergas et al. 2007; Ren et al. 2013). 
Spectral binaries are insensitive to the binary separation, 
but limited to binaries composed of stars of distinct spectral characteristics, such as 
a white dwarf and a late type star.

Binaries can also be identified indirectly and statistically. For clusters consisting of a simple stellar population, 
binary stars are brighter by up to 0.75 mag above the sequence of single stars. The magnitude spread of color-magnitude diagram
of a cluster can thus be used to constrain the binary fraction and mass ratio distribution statistically (e.g., Hu et al. 2010; Li et al. 2013). 
Very recently, by decomposing the radial velocity variations yielded by duplicate observations 
in the SDSS and LAMOST surveys into contributions from the measurement uncertainties and from generic variations from binaries, 
Gao et al. (2014) obtain an estimate of binary fraction for field FGK dwarfs of orbital periods shorter than 1,000 days.
The method is biased to close binaries and sensitive to the orbital period distribution assumed. 
Radial velocity variations of high mass ratio binaries may be not as simple as assumed  in the analysis. 

Stellar color locus defines the distribution of ``normal'' main sequence stars in the color-color space.
By combining the re-calibrated photometric (Ivezi{\'c} et al. 2007; Yuan et al. 2014a) and spectroscopic data of the SDSS Stripe 82,
Yuan et al. (2014b; hereafter Paper I) build a large, clean sample of main sequence (MS) stars with accurate colors (about 1 per cent) 
and well determined metallicities (about 0.1\,dex) to investigate the metallicity dependence and intrinsic widths of the SDSS stellar color loci.
By fitting to the $u-g$, $g-r$, $r-i$, and $i-z$ colors as a function of the $g-i$ color and \feh~with two-dimensional polynomials,
they obtained for the first time the metallicity-dependent stellar loci in the SDSS colors and 
find that the fit residuals can be fully accounted for by the uncertainties in photometric measurements, metallicity determinations and calibration,
suggesting that the intrinsic widths of the loci are at maximum of a few mmag if not zero.
More interestingly, the distributions of residuals are asymmetric, 
pointing to the presence of a significant population of binaries.
As we shall show in the current work, by modeling the distributions of residuals 
it is possible to reveal the binary fraction of field MS stars.

In this companion paper, we propose a Stellar Locus OuTlier (SLOT) method to
determine the binary fractions statistically. 
Compared to the previous techniques,
the SLOT method is solely based on color deviations relative to the metallicity-dependent stellar loci,  
and thus is independent of the separations (orbital periods) of binaries 
if one neglects the small fraction of spatially resolved wide binaries (semi-major axes $\gtrsim$ 100\,AU; Chanam{\'e} 2007).
The method is also insensitive to the assumed mass-ratio distribution.
Because binaries of intermediate mass-ratios contribute most of the observed deviations in the color-color space, while those of 
close to unity or very small mass-ratios contribute little.
With modern photometric surveys (providing accurate colors) and large scale spectroscopic surveys
such as the SDSS and LAMOST (providing robust estimates of metallicity and surface gravity),
the SLOT method is capable providing model-free estimates of the binary fraction for
large samples of stars of different populations. 
When combining results from other techniques, the method can also provide strong constraints on 
the distributions of orbital periods and mass ratios of binary stars.

In this paper, we will introduce the SLOT method, apply the method
to stars of the SDSS Stripe\,82  spectroscopically targeted by the SDSS and LAMOST.
and present a model-free estimate of the binary faction for field FGK stars.
The variations of binary fraction with spectral type, metallicity, and population are also investigated. 
The paper is organized as follows.  
The SLOT method is introduced in Section\,2. 
The data and results are presented in Section\,3.
The variations of binary fraction for different stellar populations are explored in Section\,4.
The conclusions are given in Section\,5 along with a discussion of other potential applications of the method and future work.

\section{Method}

\begin{figure}
\includegraphics[width=90mm]{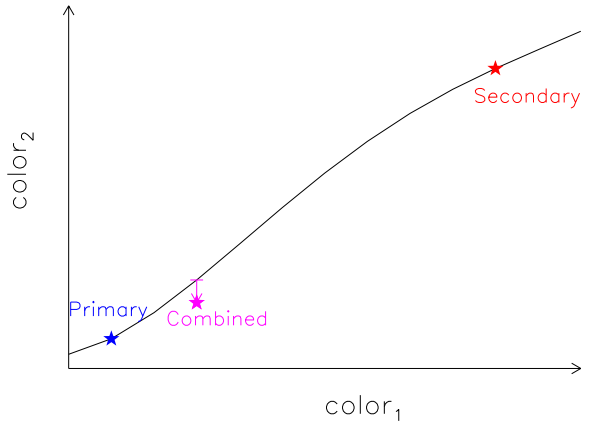}
\caption{
A plot illustrating the SLOT method. 
The line denotes the stellar locus of MS single stars of a given metallicity. 
The purle, blue, and red stars denote locations of a binary system and its primary and secondary stars, respectively. 
The colors of the binary system devirate from the stellar locus, as indicated by the arrow.
} 
\label{sketch}
\end{figure}

\begin{figure*}
\includegraphics[width=160mm]{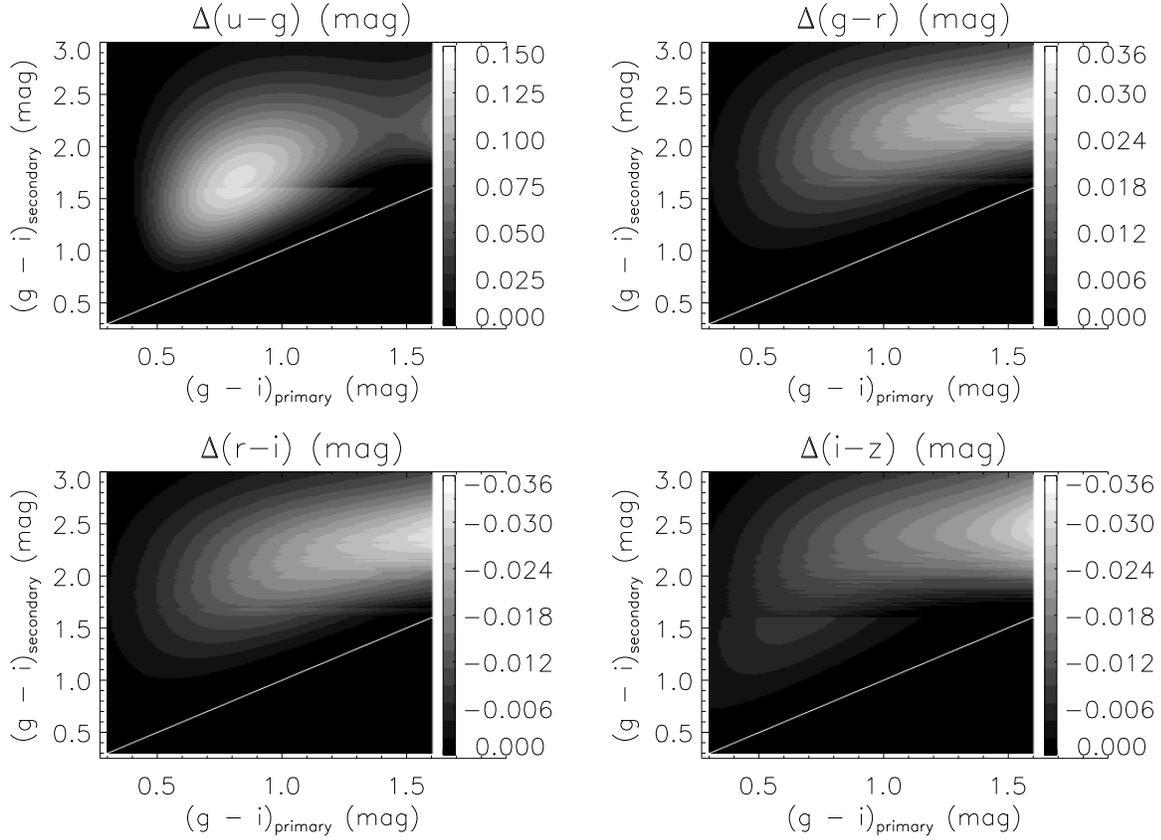}
\caption{Distributions of differences between the combined $u-g$ (top left), $g-r$ (top right), $r-i$ (bottom left), and $i-z$ (bottom right) 
colors of binary systems and those predicted by the metallicity-dependent stellar loci for the combined $g-i$ color
of the systems, as a function of the $g - i$ colors of the consisting primary and secondary stars. The binaries are assumed to 
compose of two MS single stars of \feh~=~$-$0.8\,dex. 
A colorbar is overplotted by the side in each case. The initial set of stellar loci determined in Section\,3.1.2 
of the current paper and those of Covey et al. (2007) are used for $g-i \leq 1.6$ and $>$ 1.6\,mag, respectively. 
} 
\label{effect}
\end{figure*}

\begin{figure}
\includegraphics[width=90mm]{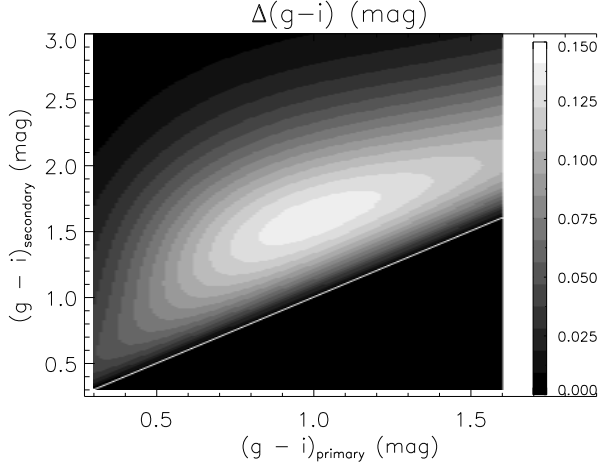}
\caption{Distributions of differences between the combined $g-i$ colors and those of the primary  
for binary systems composed of two MS single stars of \feh~=~$-$0.8\,dex
as a function of the $g-i$ colors of the primary and secondary stars.
A colorbar is overplotted. The initial set of stellar loci determined in Section\,3.1.2
of the current paper and those of Covey et al. (2007) are used for $g-i \leq 1.6$ and $>$ 1.6\,mag, respectively.
}
\label{effect_dgi}
\end{figure}

\begin{figure}
\includegraphics[width=90mm]{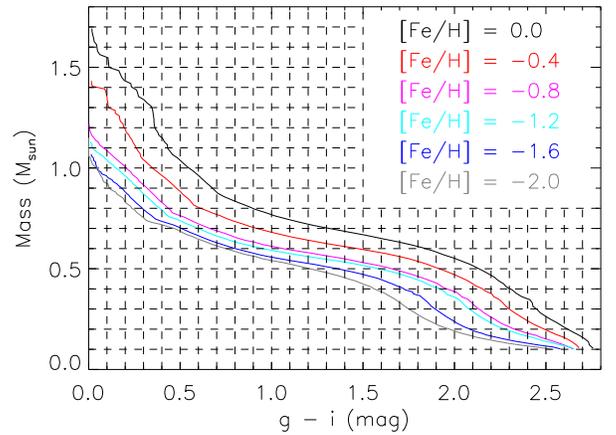}
\caption{Stellar mass as a function of $g-i$ color for MS stars at different metallicities.
The data are taken from Dotter et al. (2008).
}
\label{mass_gi}
\end{figure}

As shown in Paper\,I, the intrinsic widths of the metallicity-dependent stellar loci of the SDSS colors for MS stars are essentially zero, 
i.e., the colors $u-g$, $g-r$, $r-i$, and $i-z$ of a single MS star are fully determined by its $g-i$ color and \feh. 
To show how the SLOT method works, let consider binary systems composed of two MS single stars. 
The colors of the binary systems will deviate from what predicted by the metallicity-dependent loci (of single stars), 
as illustrated in Fig.\,\ref{sketch}.
The deviations can be simulated for all possible combinations of primary stars of $g-i$ color ranging from 0.3 -- 1.6\,mag
and secondary stars of $g-i$ color ranging from 0.3 -- 3.0\,mag, with the condition that the color of the secondary 
is no bluer than that of the primary.
For the simulation, we assume that both the primary and secondary stars have the same \feh~of $-$0.8\,dex.
For a given binary of given $g-i$ colors of the two component stars and metallicity \feh, the absolute $r$-band magnitudes, $M(r)$, 
of the two stars are 
computed using the photometric parallax relation of Ivezi{\'c} et al. [2008; Eq.\,A7], and their 
$u-g$, $g-r$, $r-i$, and $i-z$ colors are computed using the metallicity-dependent stellar loci determined 
in Section\,3.1.2 of the current paper for $g-i \leq 1.6$\,mag and
using the stellar loci of Covey et al. (2007) for $g-i > 1.6$\,mag.
The absolute $r$-band magnitudes and         
$u-g$, $g-r$, $r-i$, and $i-z$ colors are then used to derive the absolute magnitudes in $u,g,i,$ and $z$ bands.
Given the absolute magnitudes in $u,g,r,i,$ and $z$ bands of both stars, 
the combined magnitudes and colors of the binary are calculated. 
The differences between the combined $u-g$, $g-r$, $r-i$, and $i-z$ colors of the system and 
those predicted by the metallicity-dependent stellar loci (of single stars) for a combined  $g-i$ color (of the binary system)
are then deduced and plotted in Fig.\,\ref{effect} as a function of the $g - i$ colors of the primary and secondary stars, respectively.    
The deduced differences between the $g-i$ color of the system and that of the primary star are shown in Fig.\,\ref{effect_dgi}.

Fig.\,\ref{effect} shows that: 
1) Colors of binary systems generally do not follow the metallicity-dependent stellar loci of single MS stars. 
Rather they deviate from the loci in a systematic way. Colors of simulated binaries are always redder 
in $u-g$ and $g-r$ and bluer in $r-i$ and $i-z$ than values predicted for single stars. 
Such systematic deviations are actually the underlying cause of the asymmetric distributions of 
of residuals of colors with respect to the stellar loci, as reported in Paper\,I;
2) The deviations are very small. The maximum deviations are about 0.15, 0.036, $-$0.036, and $-$0.036\,mag 
in the $u-g$, $g-r$, $r-i$, and $i-z$ colors, respectively. The deviations are nearly zero 
when the primary and secondary stars are of close to unity or very small mass ratios. They reach 
the maximum values at certain colors of the primary and secondary. 
Except for $u - g$, the absolute values of maximum deviation in colors 
$g-r$, $r-i$, and $i-z$ occur at large $g-i$ colors of the primary. 
This implies that binary stars consisting of an early type primary are more easily detected in the blue $u-g$ color while 
those consisting of a late type primary are more easily detected using red colors.
Note the artifacts at the secondary $g-i$ colors of 1.6\,mag are caused 
by the different sets of stellar loci used below and above the value. 
Note also the deviations in $g-r$ and $r-i$ colors are exactly equal but opposite.
Fig.\,\ref{effect_dgi} shows that colors of binary systems composed of two MS stars 
are slightly redder by 0.0 -- 0.15\,mag than those of the consisting primary stars. 
The combined colors of such binary systems are obviously dominated by their primary stars.

Given a sample of MS stars with accurate photometric colors and 
spectroscopically determined metallicities, the metallicity-dependent 
stellar loci of the sample are obtained as in Paper\,I.
Two sets of Monte Carlo (MC) simulations are performed, 
one assumes that all stars in the sample are single while the other assumes that all stars are binaries,
in order to mimic the distributions of the observed sample in the color-color diagrams. 
For the first set of simulations, for a star of color $g-i$ and metallicity \feh, 
its $u-g$, $g-r$, and $i-z$ colors are computed as predicted by the metallicity-dependent stellar loci
if $g-i \leq 1.6$\,mag. For $g-i > 1.6$\,mag, the stellar loci of Covey et al. (2007) are used.
For the other set of simulations of binary stars, for a star of color $g-i$ and metallicity \feh, 
the mass of the primary is derived from the Dotter et al. (2008) isochrones (Fig.\,\ref{mass_gi}) 
by linear interpolation. The mass of the secondary star is then generated by MC simulation for an assumed 
mass ratio distribution of binary stars. We assume that the mass ratio distribution 
follows a power law of index $\gamma = 0.3$ for mass ratios between 0.05 and 1 (Duch{\^e}ne \& Kraus 2013).
If the mass of the secondary is smaller than 0.08\,$M_\odot$, its effects are then neglected. 
The $g-i$ color of the secondary is then derived from the same set of isochrones.
The combined colors of the binary system are then calculated as described in the beginning of this Section. 
For both sets of simulations, when calculating the predicted colors,
the effects of uncertainties of metallicity determinations and photometric measurements, 
as well as of errors of photometric calibration are fully taken into account using MC simulations. 
Considering the very small color deviations we are dealing with, 
it is essentially that all sources of errors are reliably determined and propagated in order for the SLOT method to work. 
By adjusting the relative fraction of stars in the two sets of simulations to fit the 
residual distributions with respect to the metallicity-dependent 
stellar loci of the observed sample, the binary fraction of the sample is determined.
A minimum $\chi^2$ technique is used in the fit.
Here $\chi^2$ is defined as:
\begin{equation} 
   \chi^2 = \sum_{i=1}^{M}\frac{(N_{\rm obs}^i - N_{\rm mod}^i)^2}{\sigma_i^2 \times (M-2)}, 
\end{equation} 

\begin{equation} 
   \sigma_i = (N_{\rm obs}^i + N_{\rm mod}^i)^{1/2}, 
\end{equation}

\begin{equation} 
   N_{\rm mod}^i = N_{\rm binary}^i \times f_{\rm b} + N_{\rm single}^i \times (1 - f_{\rm b}), 
\end{equation}  
where $N_{\rm obs}^i$ is the observed  numbers of stars in the $i$-th bin, 
$N_{\rm single}$ and $N_{\rm binary}$ are the predicted numbers of stars in the $i$-th bin by the 
two sets of simulations for single and binary stars, respectively, 
$N_{\rm mod}^i$ is the predicted numbers of stars in the $i$-th bin for a binary fraction $f_{\rm b}$,
$M$ is the total bin number used to calculate $\chi^2$, and $\sigma_i$
is the statistical (Poissonian) error associated with the numbers of stars $N_{\rm obs}^i$ and $N_{\rm mod}^i$. 
We vary $f_{\rm b}$ from 0.0 to 1.0 at steps of 0.01 and calculate the $\chi^2$ value for each $f_{\rm b}$.
The minimum $\chi^2$ value, $\chi^2_{\rm min}$, the associated best-fit value of $f_{\rm b}$ and one-sigma uncertainty are then determined. 
The one-sigma uncertainty corresponds to the difference of $f_{\rm b}$ values at $\chi_{\rm min}^2$ and $\chi_{\rm min}^2$+1 (Avni 1976; Wall 1996).

For small observed samples, simulations can be carried out multiple times to reduce the random errors of the simulations.
For the single-star set of simulations, the resultant distributions in $g-i$ and \feh~are essentially the same as those of the observed sample. 
For the second set of simulations for binaries, the resultant distributions in $g-i$ of the simulated sample 
may differ slightly with respect to those of the observed sample. 
To account for possible variations of the binary fraction as a function of stellar color, 
for each $g-i$ bin of 0.1 mag width, the number of stars in the second set of simulations
are adjusted to match that of the observed sample by duplicating or removing some randomly selected targets. 

The determination of binary fraction for an observed sample relies on the 
residual distributions in determining the metallicity-dependent stellar loci
for the sample. On the other hand, fitting the metallicity-dependent stellar loci 
for the sample will in no doubt affected by the possible presence of binaries in the sample.
Therefore, iterations are needed. An initial set of metallicity-dependent stellar loci are first derived by two-sigma clipping 
when fitting the data in order to reduce the effects of binary stars as in Paper\,I. With the initial set of loci, 
an estimate of the binary fraction is obtained.
The effects of binary stars are then estimated by comparing the initial set of stellar loci with those 
given by the simulations for the binary fraction derived above. The differences are corrected for. 
Note that when combining the two sets of simulations for respectively single and binary stars, 
the possible variations of the binary fraction in individual bins of stellar colors are considered. 
With the revised metallicity-dependent stellar loci and updated residual distributions that result, 
a new estimate of the binary fraction is obtained.
The above process is repeated until a convergence is achieved.
The corrections to stellar loci with respect to the initial set are very small, at a level of a few mmag. 
Given that the residual distributions with respect to the adopted metallicity-dependent stellar loci 
for the simulated samples of single or binary stars are barely affected by the small revisions 
of the metallicity-dependent stellar loci used, 
their effects on the residual distributions of the simulated samples are ignored.
The corrections only affect the residual distributions of the observed sample.

\section{Data and analysis}

\subsection{SDSS} 
\subsubsection{Data}
The SLOT method requires a sample of main sequence stars with accurate photometric colors and well determined metallicities.
The repeatedly scanned equatorial Stripe 82 ($|{\rm Dec}| < 1.266\degr$, 20$^h$34$^m$ $< {\rm RA} <$ 4$^h$00$^m$)
in the SDSS has delivered very accurate photometry
for about 1 million stars in $u,g,r,i,z$ bands (Ivezi{\'c} et al. 2007).
The data have been further calibrated by Yuan et al. (2014a) using an innovative spectroscopy-based 
Stellar Color Regression method, achieving an accuracy of about 0.005, 0.003, 0.002, and 0.002 mag
in colors $u-g$, $g-r$, $r-i$, and $i-z$, respectively.
In addition, over 40,000 stellar spectra in the region have been released in the SDSS Data Release 9 (DR9; Ahn et al. 2012),
along with the basic stellar parameters (radial velocity, effective temperature, surface gravity, and metallicity) 
deduced with the {\rm Sloan Extension for Galactic Understanding and Exploration}
({\rm SEGUE}; Yanny et al. 2009) Stellar Parameter Pipeline (SSPP; Lee et al. 2008a,b).
By combining the spectroscopic information with the re-calibrated photometry of Stripe 82,
we have constructed a large, clean sample of main sequence stars of
well determined metallicities and accurate colors and determined the 
metallicity dependent stellar loci in the SDSS colors (Paper\,I).

The sample of Paper\,I includes stars selected with a variety of specific  criteria in the SEGUE 
program\footnote{See https://www.sdss3.org/dr9/algorithms/ \\
segue\_target\_selection.php\#SEGUEts1 for details.} (Yanny et al. 2009).
Many stars in the sample are selected from the SDSS color-color diagrams 
as candidates of, for example, AGB (Asymptotic giant branch stars), KG (K giants), MP (metal-poor stars), 
MSWD (main-sequence-white-dwarf binaries), and QSO. 
The specific selection algorithms may thus cause over- or under-selection of binary stars.
To avoid such effects, targets selected based on the color-color diagrams are excluded. 
The sample of the current work is the same as Paper\,I except that only stars selected 
as candidates of BHB (blue horizontal branch stars), FG (F or G dwarfs), 
GD (G dwarfs), HHV (halo high velocity stars), HOT (hot standard stars), HVS (hypervelocity stars), LKG (low latitude K giants), 
MKD (M or K dwarfs), PHO (photometric standard stars), and RED (reddening standard stars) are included.
Note that due its possilbe dependece on spectral types and metallicities,  
the binary fraction yielded by a given sample of stars, depending on the distributions of the sample in color and metallicity, 
may not reflect the characteristics of the whole Milky Way stellar population.

In total, 14,650 stars are selected, most are candidates of BHB, GD, MKD, PHO, and RED.
Their distribution in the $g-i$ and \feh~plane is shown in Fig.\,5. 
Note that in this work all colors refer to the dereddened values. 
The reddening corrections are performed using the extinction map of Schlegel et al. (1998) and the empirical 
reddening coefficients of Yuan et al. (2014a), derived using the star pair technique (Yuan, Liu \& Xiang 2013).
The average spectral signal-to-noise ratios (SNRs) of the sample stars plotted against 
the r-band magnitudes and the photometric errors of the individual bands plotted against the 
magnitudes of the corresponding band are shown in Fig.\,6.
The photometric errors in $u$ band increase rapidly with the $u$ magnitudes. 
The errors are about 0.01 mag at $u=19.0$\,mag and 0.1 mag at $u=22$\,mag.
The photometric errors in $g,r,i$ bands are dominated by the calibration uncertainties and essentially constant, 
at the level of $0.006\pm0.001$, $0.005\pm0.001$, and $0.005\pm0.001$ mag, respectively.
The photometric errors in $z$ band increase modestly with the $z$ magnitudes, 
0.01 mag at $z=18.2$\,mag and 0.02\,mag at $z=19$\,mag. A test of error estimates by Ivezi{\'c} et al. (2007) shows that 
the magnitude errors computed by the photometric pipeline are reliable for the $g,r,i$, and $z$ bands.
For the $u$ band, the errors may have been slightly underestimated, by about 10 per cent. 
As a consequence, we have multiplied the photometric errors in the $u$ band by 1.1 in the current work.  

As described in the previous Section, the SLOT method also requires robust estimates of errors of \feh.
To estimate the random errors of \feh~yielded by the SSPP pipeline,
13,270 duplicate observations of comparable spectral SNRs of stars that fall in the 
parameter ranges of the current sample
(4,300 $\le$ \teff~$\le$ 7,000\,K, \logg~$\ge$ 3.5\,dex, $-2.0$ $\le$ \feh~$\le$ 0.0\,dex)
are selected from the SDSS DR9 samples.
Given the relatively narrow range of effective temperature of the sample stars, 
the random errors of \feh~ are fitted as a function of SNR and \feh~of the following form: 
\begin{align}
\sigma_{\rm {ran}}({\rm [Fe/H]}) = \nonumber\\ 
& a_0 +  a_1 \times {\rm [Fe/H]} + a_2 \times ({\rm [Fe/H]})^2 +\nonumber\\ 
&a_3 \times {\rm SNR} + a_4 \times {\rm SNR} \times {\rm [Fe/H]} + \nonumber\\ 
&a_5 \times {\rm SNR}^2.
\end{align}
Only stars of SNRs between 10 -- 50 are used in the fitting.
The resultant fit coefficients $a_0$ -- $a_5$ are 0.17, $-$0.066, $-$0.0088, $-$0.0063, $7.0\times 10^{-4}$, and $6.8\times 10^{-5}$, respectively.
When assigning random errors of \feh~for stars of SNRs higher than 50, the values given by the above equation for a SNR of 50 are used.
For SNR = 10, the random errors of \feh~are about 0.11, 0.16, and 0.19\,dex for \feh~= 0, $-$1, and $-$2\,dex, respectively.
For SNR = 50, the errors are 0.02, 0.04, and 0.05\,dex for \feh~= 0, $-$1, and $-$2\,dex, respectively.
For the systematic errors, we have simply adopted the relation:
\begin{equation} 
  \sigma_{\rm {sys}}({\rm [Fe/H]}) = 0.03 - 0.05 \times {\rm [Fe/H]}, 
\end{equation}
based on the analysis of the dispersions of \feh~of member stars of clusters (Lee et al. 2008b; Smolinski et al. 2011), 
where \feh~refers to the SSPP adopted values.
Note that \feh~values yielded by the SSPP may be affected by the binarity, i.e., the presence of a secondary star.  
However, the effects are likely to be insignificant in most case (Schlesinger et al. 2010).
The maximum additional uncertainty caused by binary contamination at SNR = 10 is about 0.17\,dex in \feh~(Schlesinger et al. 2010).
The uncertainties of \feh~error estimates affect mainly the $u-g$ color.
Their effects on the $g-r$ and $i-z$ colors are negligible.

\begin{figure}
\includegraphics[width=90mm]{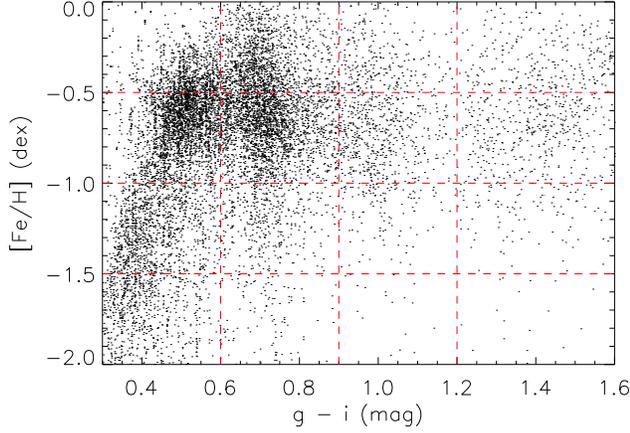}
\caption{
Distribution of the selected SDSS DR9 stellar spectroscopic sample of Stripe 82  in the $g - i$ and [Fe/H] plane.
}
\label{}
\end{figure}

\begin{figure}
\includegraphics[width=90mm]{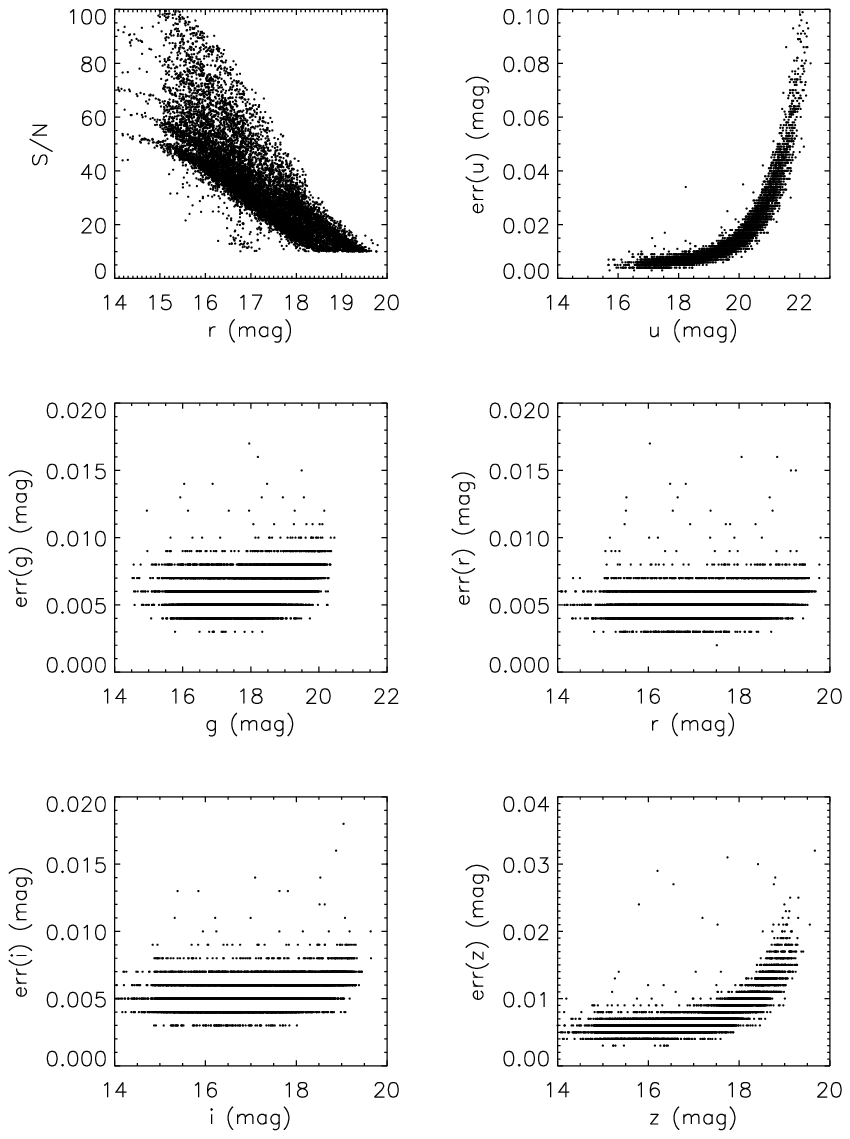}
\caption{The average spectral SNRs plotted against the $r$-band magnitudes 
for the selected SDSS DR9 stellar spectroscopic sample of Stripe 82 (top left panel).
The remaining panels show the photometric errors of the individual bands plotted against the magnitudes of the corresponding band.
}
\label{}
\end{figure}

\subsubsection{Results}
Using the sample selected above, we have carried out a global two dimensional polynomial fit to 
the $u-g$, $g-r$, $r-i$, and $i-z$ colors as a function of color $g-i$ and metallicity \feh~to 
determine a initial set of stellar color loci.
As in Paper\,I, a 4th-order polynomial with 15 free parameters is adopted for color $u-g$
and a 3rd-order polynomial of 10 free parameters is used for the other three colors.
Two-sigma clipping is performed during the fitting process.
The resultant fit coefficients are listed in the upper part of Table\,1.
Fig.\,7 compares this initial set of stellar color loci obtained here with those of Paper\,I.
The differences are less than 0.03\,mag in $u-g$ and 0.01\,mag in other colors.

\begin{table} 
\centering
\caption{Fit coefficients of the stellar color loci for the SDSS sample.}
\label{}
\begin{tabular}{lrrrr} \hline\hline
Coeff. & $u-g^{a)}$ & $g-r^{b)}$ & $r-i^{b)}$ & $i-z^{b)}$  \\\hline
\noalign{\vskip2pt} \multicolumn{5}{c}{Initial set}  \\
$a_0$ &   1.5862 &   0.0548 &  $-$0.0548 &  $-$0.0806 \\
$a_1$ &   0.2102 &   0.0313 &  $-$0.0313 &  $-$0.0116 \\
$a_2$ &   0.4032 &   0.0208 &  $-$0.0208 &  $-$0.0059 \\
$a_3$ &   0.2020 &   0.0039 &  $-$0.0039 &  $-$0.0024 \\
$a_4$ &   0.0356 &   0.6244 &   0.3756 &   0.1780 \\
$a_5$ &  $-$3.5203 &   0.0324 &  $-$0.0324 &  $-$0.0494 \\
$a_6$ &   0.7480 &  $-$0.0005 &   0.0005 &  $-$0.0170 \\
$a_7$ &  $-$0.1826 &   0.1329 &  $-$0.1329 &   0.0322 \\
$a_8$ &  $-$0.0711 &  $-$0.0158 &   0.0158 &   0.0092 \\
$a_9$ &   8.3834 &  $-$0.0546 &   0.0546 &  $-$0.0172 \\
$a_{10}$ &  $-$0.5626 &    &    &    \\
$a_{11}$ &  $-$0.0070 &    &    &    \\
$a_{12}$ &  $-$5.5546 &    &    &    \\
$a_{13}$ &   0.0606 &    &    &    \\
$a_{14}$ &   1.2053 &    &    &    \\\hline
\noalign{\vskip2pt} \multicolumn{5}{c}{Final set}  \\
$a_0$ &   1.6058 &   0.0439 &  $-$0.0439 &  $-$0.0754 \\
$a_1$ &   0.1415 &   0.0272 &  $-$0.0272 &  $-$0.0070 \\
$a_2$ &   0.3637 &   0.0211 &  $-$0.0211 &  $-$0.0056 \\
$a_3$ &   0.1873 &   0.0041 &  $-$0.0041 &  $-$0.0025 \\
$a_4$ &   0.0331 &   0.6611 &   0.3389 &   0.1575 \\
$a_5$ &  $-$3.8224 &   0.0447 &  $-$0.0447 &  $-$0.0638 \\
$a_6$ &   0.9185 &  $-$0.0004 &   0.0004 &  $-$0.0187 \\
$a_7$ &  $-$0.1661 &   0.1064 &  $-$0.1064 &   0.0430 \\
$a_8$ &  $-$0.0704 &  $-$0.0224 &   0.0224 &   0.0152 \\
$a_9$ &   9.2585 &  $-$0.0496 &   0.0496 &  $-$0.0178 \\
$a_{9}$ &   9.2585 &  $-$0.0496 &   0.0496 &  $-$0.0178 \\
$a_{10}$ &  $-$0.7458 &   &    &    \\
$a_{11}$ &  $-$0.0112 &   &    &    \\
$a_{12}$ &  $-$6.3652 &    &   &    \\
$a_{13}$ &   0.1232 &    &   &    \\
$a_{14}$ &   1.4432 &    &   &    \\
\hline
\end{tabular}
\begin{description}
\item[$^{a)}$]  $f(x,y)=a_0+a_1*y+a_2*y^2+a_3*y^3+a_4*y^4+a_5*x+$ \\
             $ a_6*x*y+a_7*x*y^2+a_8*x*y^3+a_9*x^2+a_{10}*y*x^2+$ \\
             $a_{11}*y^2*x^2+a_{12}*x^3+a_{13}*y*x^3+a_{14}*x^4$, where $x$ denotes color $g-i$ and $y$ denotes metallicity \feh.
\item[$^{b)}$]  $f(x,y)=a_0+a_1*y+a_2*y^2+a_3*y^3+a_4*x+a_5*x*y+a_6*x*y^2+ $ \\
             $a_7*x^2+a_8*y*x^2+a_9*x^3$ , where $x$ denotes color $g-i$ and $y$ denotes metallicity \feh.
\end{description} 
\end{table}

In order to explore the binary fraction of field stars as a function of metallicity, 
the sample has been divided into four bins of metallicity: $-2.0 \le$ \feh~$< -1.5$\,dex,
$-1.5 \le$ \feh~$< -1.0$\,dex, $-1.0 \le$ \feh~$< -0.5$\,dex, and $-0.5 \le$ \feh~$\le$ 0.0\,dex.
The fit residuals in colors $u-g$, $g-r$, and $i-z$ as a function of $g-i$ for the SDSS whole- and 
sub-samples of different metallicity ranges are shown in Fig.\,8. The fit residuals of 
color $r-i$ are not shown because they degenerate with those of color $g-r$.
As already pointed out in Paper\,I, there is an excess of stars whose colors, 
when compared to what predicted by the fits, are redder in $u-g$ and $g-r$ but bluer in $i-z$.
The asymmetries are most prominent for stars of red colors. The asymmetric residuals are fully consistent with 
the presence of binary stars in the sample in terms of the ranges and directions of the offsets as shown in Fig.\,2.

For the SDSS sample, two sets of MC simulations are performed:
one assumes that all sample stars are single while the other assumes that all sample stars are binaries.
The detailed procedures have been described in the previous Section.
To reduce the random errors of the simulations, each set of simulations is carried out ten times and then combined.
Fig.\,9 shows the offsets of colors $u-g$, $g-r$, and $i-z$ relative to those predicted by the metallicity-dependent stellar loci 
as a function of $g-i$ for the simulated whole- and sub-samples of different metallicity ranges. 
Only one in ten randomly selected targets are shown. 
By adjusting the fractions of stars in the two sets of simulations to fit the observed residual distributions 
with respect to the metallicity-dependent stellar loci, 
the binary fractions of the SDSS whole and sub-samples can be determined. 
In this process, the whole- and sub-samples of different metallicity bins 
are further divided into bins of $g-i$ color.
The binary fractions of individual bins of color and metallicity are determined in order to 
investigate the possible dependence of binary fraction on spectral type and metallicity.
As described in the previous Section, several iterations are performed to obtain the final set of stellar loci 
devoid of contamination of binary stars in the observed sample and convergence of estimate of the binary fraction.
The final set of stellar loci derived from the SDSS sample are plotted in Fig.\,7. 
Compared to the initial ones, the differences in colors $u-g$, $g-r$, $r-i$, and $i-z$ are 
respectively within the ranges of $-30$  -- 25.4, 0.0 -- 3.5, $-3.5$ -- 0.0, and 
$-3.4$ -- $-0.4$\,mag at \feh~=~$-0.5$\,dex, and 
respectively within the ranges of $-7.8$  -- 21.6, 0.0 -- 2.1, $-2.1$ -- 0.0, and
$-2.3$ -- $-1.0$\,mmag at \feh~=~$-1.5$\,dex.
Considering the possible maximum offsets produced by the binary stars in the sample or caused by 
various potential sources of error, in the current work the fit ranges of residuals for colors 
$u-g$, $g-r$, and $i-z$ in the current work are set at respectively $-0.1$ -- 0.2, $-0.03$ -- 0.045, 
and $-0.05$ -- 0.02\,mag, with bin sizes of 0.01, 0.0025, and 0.005\,mag, respectively.

\begin{figure}
\includegraphics[width=90mm]{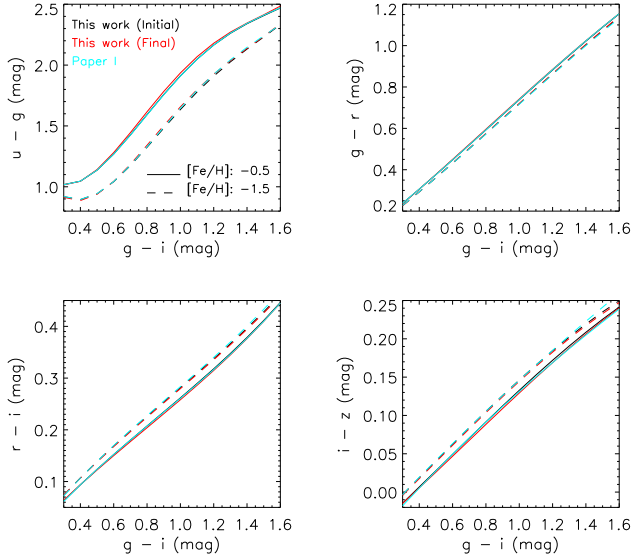}
\caption{
Comparisons of SDSS stellar color loci at \feh~=~$-$0.5\,dex (solid lines) and $-$1.5\,dex (dashed lines) 
obtained in the current work and those deduced in Paper\,I. 
}
\label{}
\end{figure}

\begin{figure*}
\includegraphics[width=180mm]{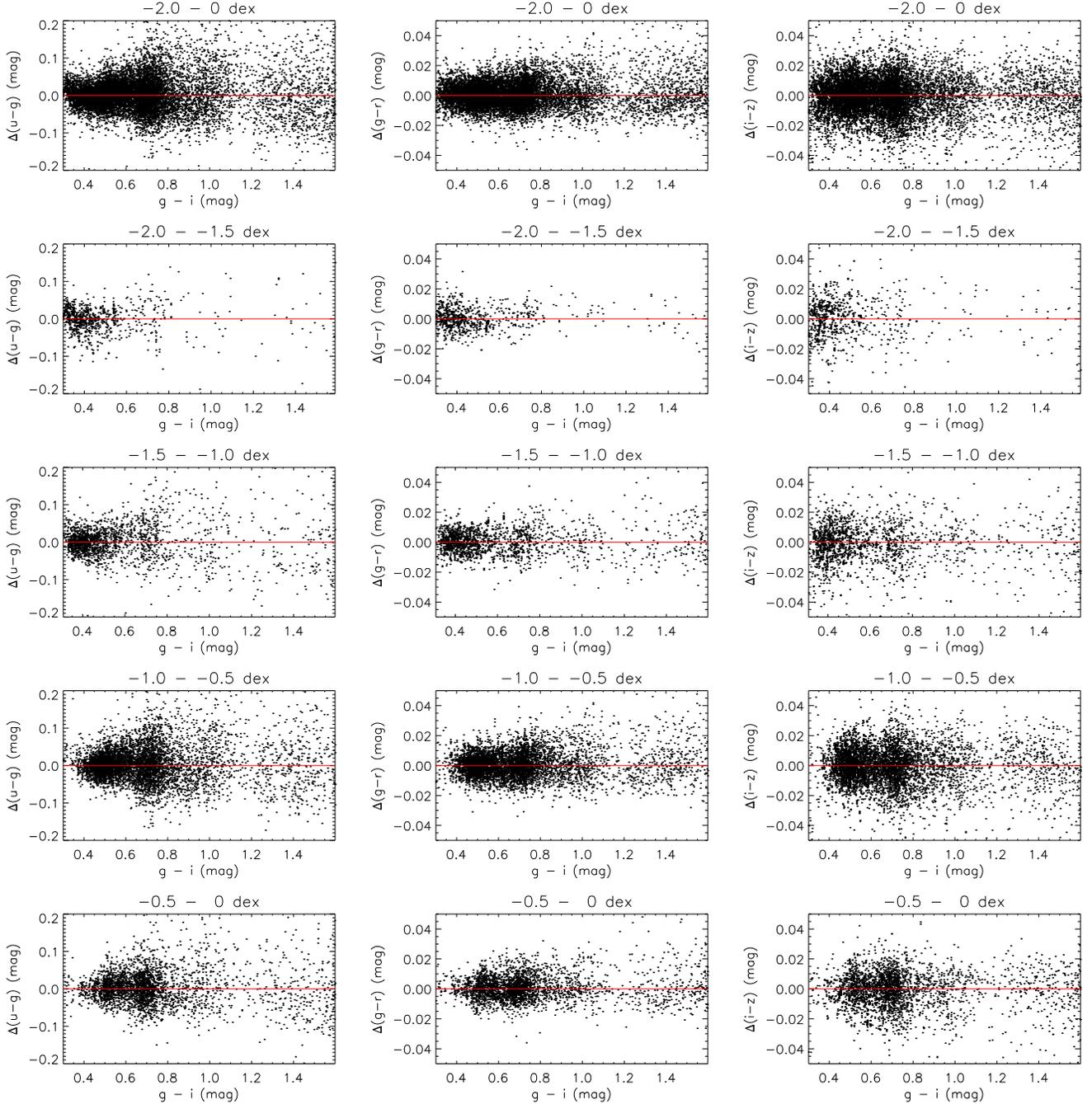}
\caption{
From top to bottom, fit residuals of colors $u-g$ (left), $g-r$ (middle), and $i-z$ (right) as a function of $g-i$ for the 
SDSS whole- and sub-samples of different metallicity ranges. 
The metallicity ranges are labeled on the top of each panel. The horizontal lines denote zero residuals.}
\label{}
\end{figure*}

\begin{figure*}
\includegraphics[width=180mm]{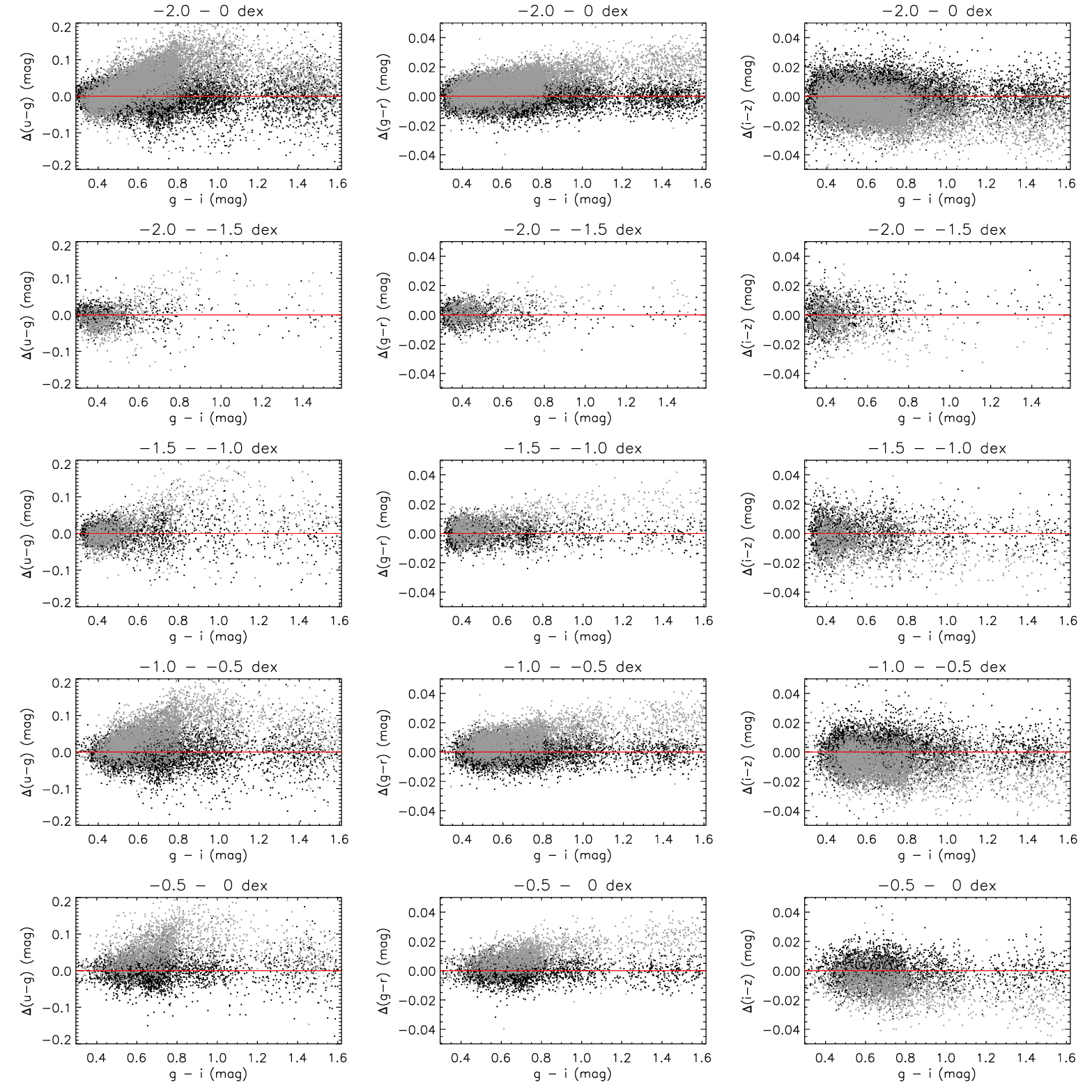}
\caption{
Same as Fig.\,8 but  
for the simulated SDSS whole- and sub-samples of single (black dots) and binary stars (grey dots).
To avoid crowdness, only one in ten stars are plotted.
}
\label{}
\end{figure*}

Figs.\,10, 11, and 12 show the distributions of the observed residuals and of the best fit models for the SDSS whole- and sub-samples 
of individual bins of $g - i$ color and metallicity, for colors $u-g$, $g-r$, and $r-i$, respectively. 
The ranges of color and metallicity of the bin, the resultant binary fraction and error, and the associated value
of minimum $\chi^2$ are marked for each panel. 
The fits are reasonably good. The typical minimum $\chi^2$ values are between 0.8 -- 3.0, 
0.8 -- 2.0 and 1.0 -- 5.0 for colors $u-g$, $g-r$ and $r-i$, respectively.
The binary fractions derived from the data of individual colors, are listed in Table\,2 and plotted in Fig.\,13
as a function of color for the individual bins of metallicity.
The weighted means by combing the results from all three colors are also listed and plotted, 
with weights determined by the corresponding uncertainties.
Considering the smaller photometric errors of $g, r$, and $i$ bands compared to those of $u$ and $z$ bands, 
the weaker dependence of color $g-r$ on metallicity compared to that of $u-g$, 
and the better residual fits in color $g-r$  compared to those in $u-g$ and $i-z$, 
we have increased the weights of results yielded by color $g-r$ by a factor of two.

The results from the individual colors agree well. For example, the binary fractions for 
the whole sample inferred from colors $u-g$, $g-r$, and $r-i$ are respectively 
$40\%\pm5\%$, $40\%\pm5\%$, and $43\%\pm4\%$, leading to a weighted mean of $42\%\pm2\%$.
The results are consistent with those determined for stars in the solar neighborhood 
(e.g., Raghavan et al. 2010). The current study also shows that the binary fraction of field FGK dwarfs 
decreases towards redder colors. The corresponding binary fractions 
are $44\%\pm5\%$, $43\%\pm3\%$, $35\%\pm5\%$, and $28\%\pm6\%$, 
for stars of $g-i$ colors between 0.3 -- 0.6, 0.6 -- 0.9, 0.9 -- 1.2, and 1.2 -- 1.6\,mag, respectively. 
The trend is consistent with the finding of previous studies for FGK stars that 
blue, massive stars seem to more often have a companion than red, less massive ones
(e.g., Eggleton \& Tokovinin 2008; Raghavan et al. 2010; Gao et al. 2014).   
The trend is also consistent with the overall trend that the binary fraction decreases 
from the early type OBA stars (e.g., Mason et al. 1998, 2009; Kobulnicky \& Fryer 2007; Kouwenhoven et al. 2007)
to the very late M-dwarfs and brown dwarfs (e.g., 
Allen et al. 2007; Burgasser et al. 2003; Fischer \& Marcy 1992; Henry \& McCarthy 1990; Joergens 2008; 
Maxted et al. 2008; Siegler et al. 2005).
This observed trend of increasing binary fraction with increasing primary mass 
is well reproduced by radiative hydrodynamical simulations (Bate 2014).

The binary fraction of field FGK stars is also found to increase with decreasing metallicity.
For stars of \feh~between $-0.5$ -- 0.0, $-1.0$ -- $-0.5$, $-1.5$ -- $-1.0$, and $-2.0$ -- $-1.5$\,dex, 
the inferred binary fractions are $37\%\pm3\%$, $39\%\pm3\%$, $50\%\pm9\%$, and $53\%\pm20\%$, respectively.
The trend is likely stronger for red than blue stars.
In spite of the very small fraction of metal-poor stars in the solar neighborhood, a number of studies 
have investigated the possible dependence of binary fractions on metallicity. 
Some studies find a lower binary fraction for subdwarfs when compared to corresponding MS stars 
(e.g., Riaz et al. 2008; Jao et al. 2009; Lodieu et al. 2009; Rastegaev 2010).
However, some of those results are based on samples biased against binary stars.
Yet some other studies find no obvious dependence of the binary fraction on metallicity (e.g., Latham et al. 2002; Chanam{\'e} \& Gould 2004).
Our results, based on a large stellar sample and an innovative, model-free method, 
agree with those found by Grether \& Lineweaver (2007), Raghavan et al. (2010), and Gao et al. (2014) that metal-poor 
stars are more likely to have a companion than their mental-rich counterpart.
If the trend is true, it may imply that lower-metallicity clouds are more liable to fragmentation  
and the formation of binary stars, as have been suggested by some numerical simulations 
(e.g., Machida et al. 2009). Alternatively, 
the initial binary fractions of stars formed under different metallicities are the same, 
but stars of different Galactic populations
(e.g., the disk and halo populations) have evolved differently because of the 
different environments that eventually lead to the observed trend.

\begin{figure*}
\includegraphics[width=180mm]{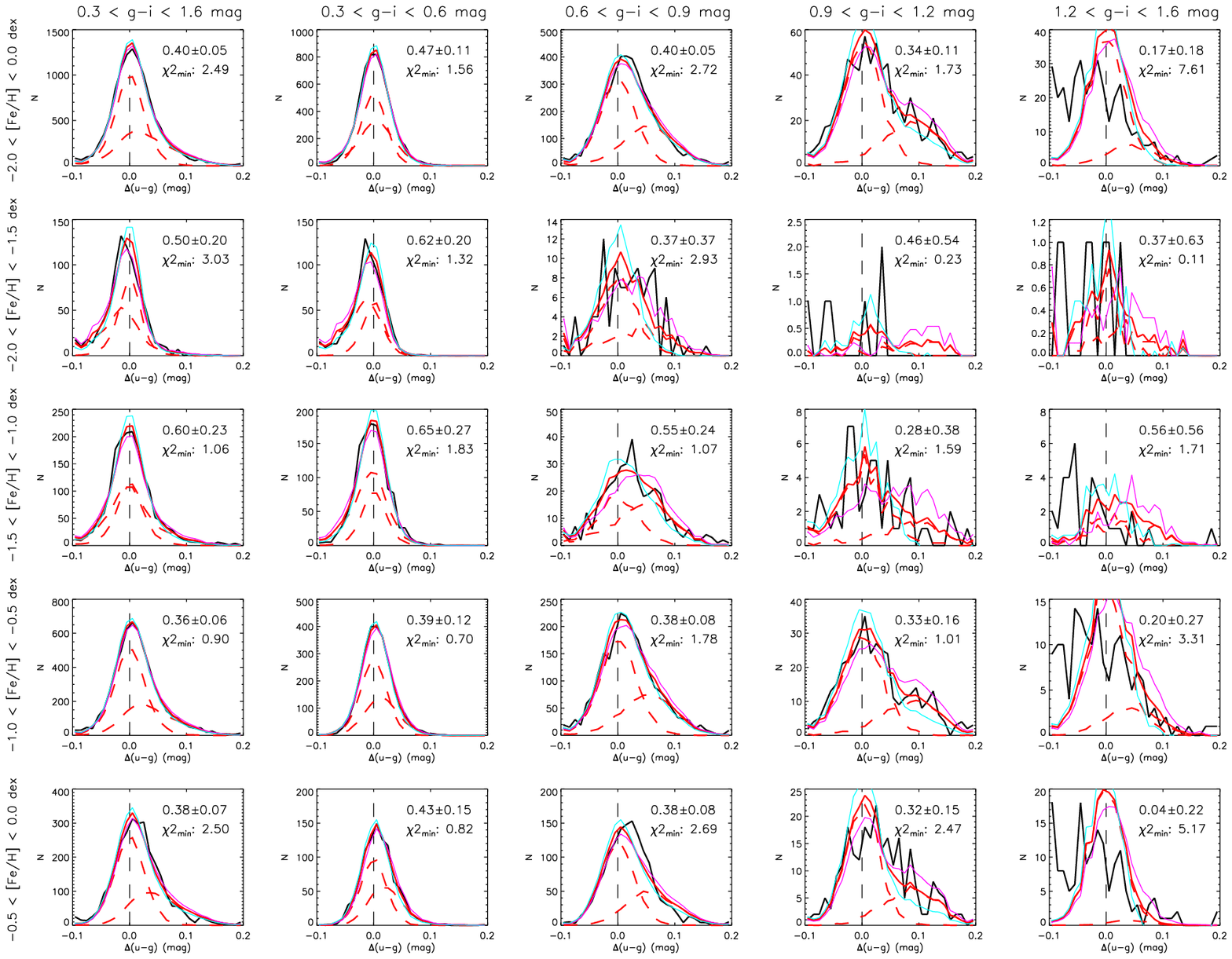}
\caption{
Distributions of the observed residuals (black lines) in $u-g$ color and the best fit models (red solid lines) 
for the SDSS sample of individual bins of color and metallicity.
The ranges of color and metallicity for each column and row are marked on the top and to the left of the figure, respectively. 
The two red dashed lines in each panel denote the contributions from the two simulated samples, 
of single and binary stars, respectively. 
The resultant binary faction, error, and the corresponding value of minimum $\chi^2$ are labeled in each panel.
The cyan and purple lines denote the fits plus and minus 1-sigma uncertainties. 
The vertical dashed lines represent zero residuals.
}
\label{}
\end{figure*}

\begin{figure*}
\includegraphics[width=180mm]{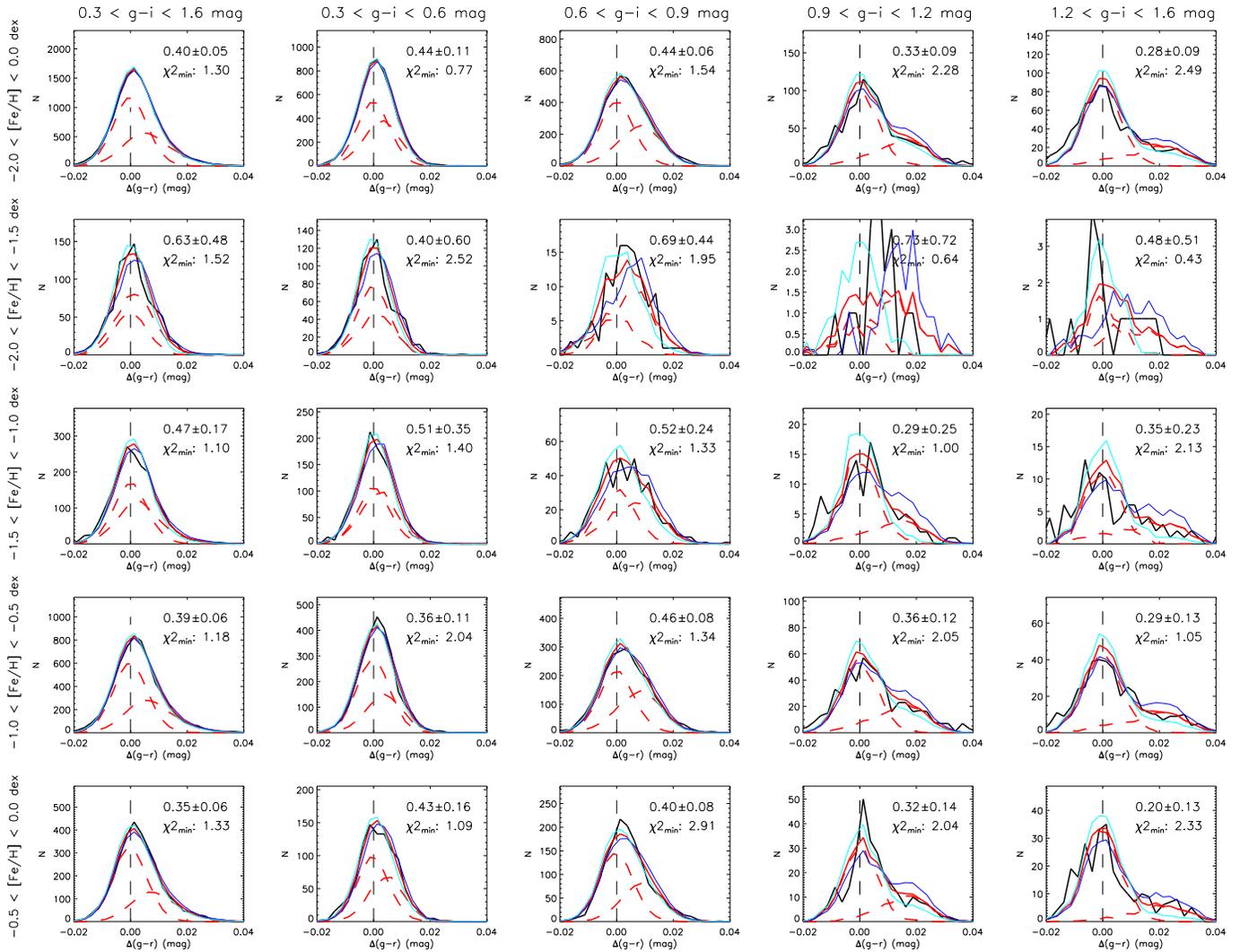}
\caption{
Same as Fig.\,10 but for the $g-r$ color.
}
\label{}
\end{figure*}

\begin{figure*}
\includegraphics[width=180mm]{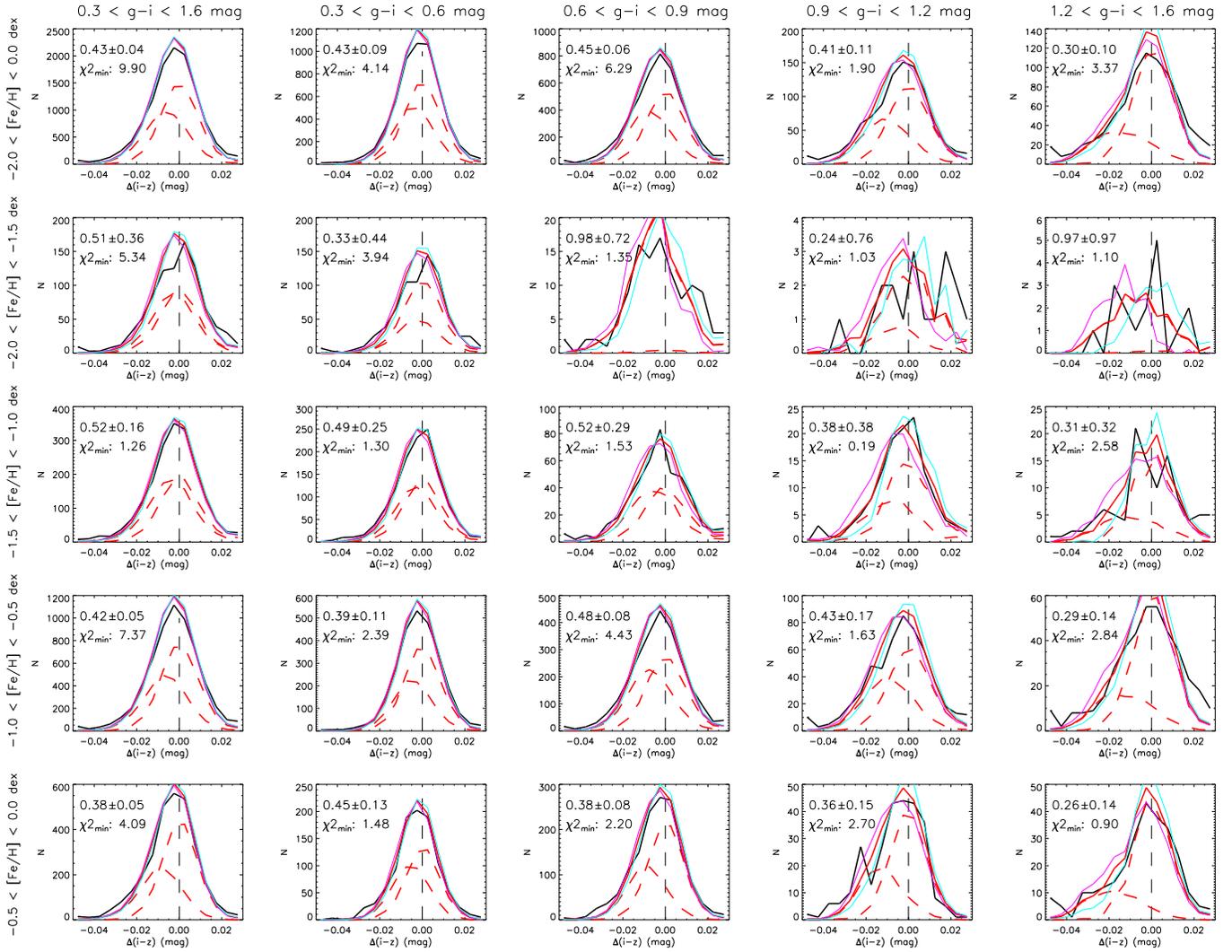}
\caption{
Same as Fig.\,10 but for the $i-z$ color.
}
\label{}
\end{figure*}

\begin{figure*}\centering
\includegraphics[width=150mm]{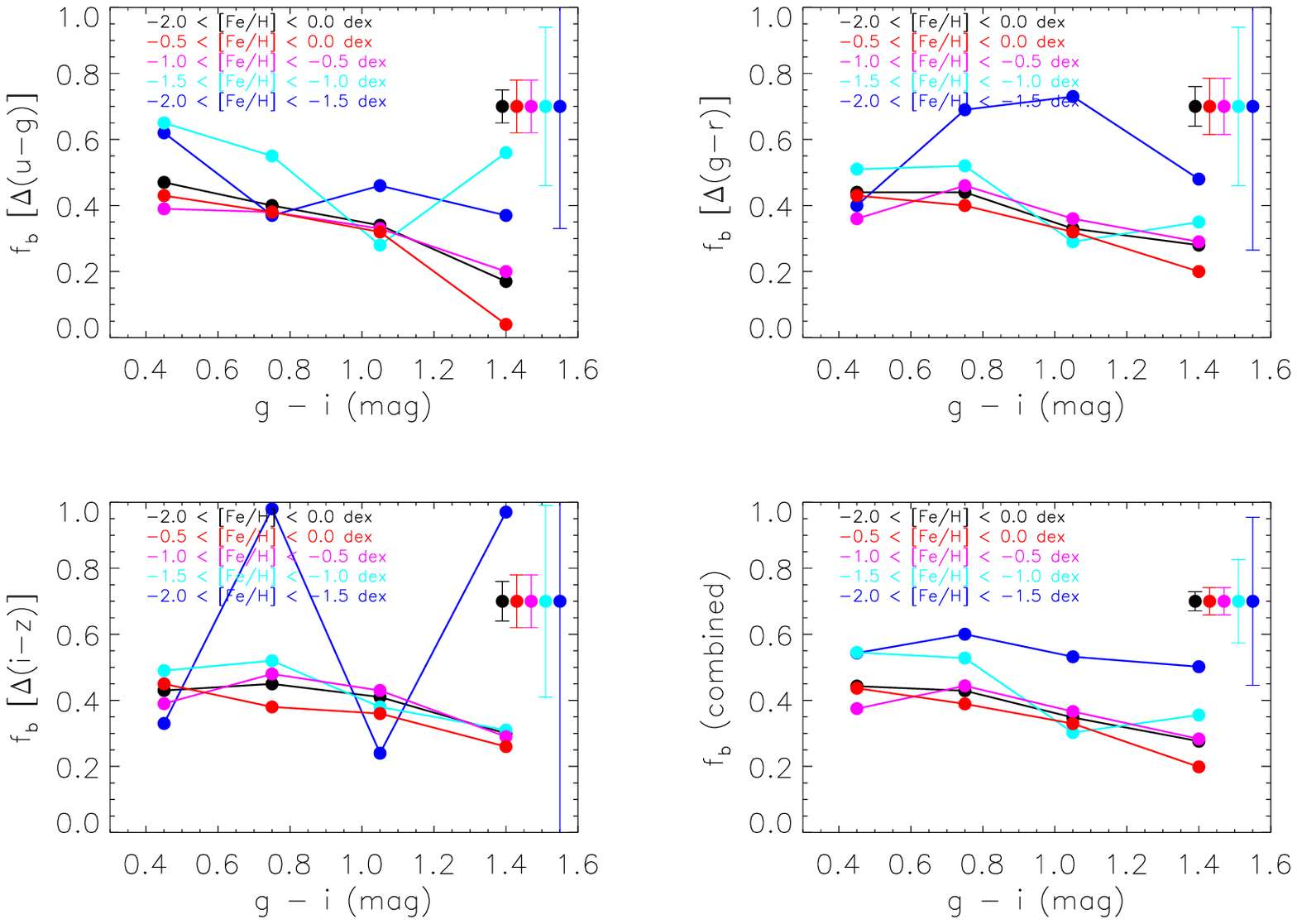}
\caption{
Binary fractions derived from residuals of the colors $u-g$ (top left), $g-r$ (top right), 
$i-z$ (bottom left) and from the combined data of all three colors (bottom right) for field FGK stars of the SDSS sample, 
plotted against $g-i$ color for the individual bins of color and metallicity. 
The typical error-bars are marked in the top-right corner of each panel. 
}
\label{}
\end{figure*}

\begin{table*} 
\centering
\caption{Binary fractions of field FGK stars of the SDSS sample.} 
\label{}
\begin{tabular}{lccccc} \hline\hline
    &  $0.3 \le g-i \le 1.6$ &  $0.3 \le g-i < 0.6$ & $0.6 \le g-i < 0.9$ & $0.9 \le g-i < 1.2$ & $1.2 \le g-i \le 1.6$  \\
    &  (mag) & (mag) & (mag) & (mag) & (mag) \\
\noalign{\smallskip}\hline  \noalign{\vskip2pt} \multicolumn{6}{c}{Based on color $u-g$}   \\
$-2.00 \le \feh < -1.50$\,dex & $0.49\pm0.21$ & $0.61\pm0.20$ & $0.35\pm0.37$ & $0.40\pm0.60$ & $0.35\pm0.65$ \\
$-1.50 \le \feh < -1.00$\,dex & $0.59\pm0.23$ & $0.65\pm0.27$ & $0.55\pm0.25$ & $0.28\pm0.38$ & $0.56\pm0.56$ \\
$-1.00 \le \feh < -0.50$\,dex & $0.36\pm0.06$ & $0.39\pm0.12$ & $0.38\pm0.08$ & $0.33\pm0.16$ & $0.20\pm0.29$ \\
$-0.50 \le \feh \le  0.00$\,dex & $0.38\pm0.07$ & $0.43\pm0.15$ & $0.38\pm0.08$ & $0.32\pm0.15$ & $0.04\pm0.24$ \\
$ -2.00 \le \feh \le 0.00$\,dex & $0.40\pm0.05$ & $0.47\pm0.11$ & $0.40\pm0.06$ & $0.34\pm0.11$ & $0.18\pm0.18$ \\
\noalign{\smallskip}\hline  \noalign{\vskip2pt} \multicolumn{6}{c}{Based on color $g-r$ }   \\
$-2.00 \le \feh < -1.50$\,dex & $0.63\pm0.48$ & $0.41\pm0.58$ & $0.69\pm0.44$ & $0.73\pm0.72$ & $0.48\pm0.51$ \\
$-1.50 \le \feh < -1.00$\,dex & $0.47\pm0.17$ & $0.51\pm0.35$ & $0.52\pm0.24$ & $0.28\pm0.25$ & $0.35\pm0.24$ \\
$-1.00 \le \feh < -0.50$\,dex & $0.39\pm0.06$ & $0.36\pm0.11$ & $0.46\pm0.08$ & $0.36\pm0.12$ & $0.29\pm0.13$ \\
$-0.50 \le \feh \le 0.00$\,dex & $0.35\pm0.06$ & $0.43\pm0.16$ & $0.40\pm0.08$ & $0.32\pm0.14$ & $0.20\pm0.13$ \\
$ -2.00 \le \feh \le 0.00$\,dex & $0.40\pm0.05$ & $0.44\pm0.11$ & $0.44\pm0.06$ & $0.33\pm0.09$ & $0.28\pm0.09$ \\
\noalign{\smallskip}\hline  \noalign{\vskip2pt} \multicolumn{6}{c}{Based on color $i-z$ }   \\
$-2.00 \le \feh < -1.50$\,dex & $0.52\pm0.35$ & $0.34\pm0.44$ & $0.99\pm0.75$ & $0.24\pm0.76$ & $0.97\pm0.97$ \\
$-1.50 \le \feh < -1.00$\,dex & $0.52\pm0.16$ & $0.49\pm0.25$ & $0.52\pm0.30$ & $0.38\pm0.38$ & $0.31\pm0.33$ \\
$-1.00 \le \feh < -0.50$\,dex & $0.42\pm0.05$ & $0.39\pm0.11$ & $0.48\pm0.08$ & $0.43\pm0.17$ & $0.29\pm0.14$ \\
$-0.50 \le \feh \le 0.00$\,dex & $0.38\pm0.05$ & $0.45\pm0.13$ & $0.38\pm0.08$ & $0.36\pm0.15$ & $0.26\pm0.15$ \\
$ -2.00 \le \feh \le 0.00$\,dex & $0.43\pm0.04$ & $0.43\pm0.09$ & $0.45\pm0.06$ & $0.41\pm0.11$ & $0.30\pm0.10$ \\
\noalign{\smallskip}\hline  \noalign{\vskip2pt} \multicolumn{6}{c}{Combined }   \\
$-2.00 \le \feh < -1.50$\,dex & $0.53\pm0.20$ & $0.54\pm0.24$ & $0.59\pm0.26$ & $0.52\pm0.35$ & $0.50\pm0.34$ \\
$-1.50 \le \feh < -1.00$\,dex & $0.50\pm0.09$ & $0.55\pm0.16$ & $0.53\pm0.13$ & $0.30\pm0.16$ & $0.36\pm0.18$ \\
$-1.00 \le \feh < -0.50$\,dex & $0.39\pm0.03$ & $0.38\pm0.06$ & $0.44\pm0.04$ & $0.37\pm0.07$ & $0.28\pm0.09$ \\
$-0.50 \le \feh \le 0.00$\,dex& $0.37\pm0.03$ & $0.44\pm0.08$ & $0.39\pm0.04$ & $0.33\pm0.07$ & $0.20\pm0.08$ \\
$ -2.00 \le \feh \le 0.00$\,dex & $0.41\pm0.02$ & $0.44\pm0.05$ & $0.43\pm0.03$ & $0.35\pm0.05$ & $0.28\pm0.06$ \\
\hline
\end{tabular}
\end{table*}

\clearpage

\subsection{LAMOST}
\subsubsection{Data}
There are 4,580 stars in Stripe 82 targeted by the LAMOST and released in the LAMOST DR1 (Luo et al. 2012; Bai et al. 2014).
Their basic stellar parameters have been determined with the LAMOST stellar parameter pipeline (LASP; Wu et al. 2014).
The LAMOST Galactic surveys (Deng et al. 2012; Liu et al. 2014) target stars randomly selected on the 
color-magnitude diagrams (Carlin et al. 2012; Yuan et al. 2014c). 
Consequently, all those stars can be used for binary fraction determinations.
By combining the LAMOST spectroscopic information and the re-calibrated SDSS photometry of Stripe 82,
we have constructed a sample of 3,827 stars of
well determined metallicities and accurate colors from the LAMOST DR1.
The selection criteria are similar to those used for the SDSS sample except that 
we require a spectral SNR $>$ 6 and $-1.0 \le \feh~\le 0.5$\,dex.
Their distribution in the $g-i$ and \feh~plane is shown in Fig.\,14.
The spectral SNRs of the sample stars plotted against
the r-band magnitudes, and the photometric errors of the individual bands plotted against the
magnitudes of the corresponding band are shown in Fig.\,15.
Due to the relatively low spectral SNRs of the LAMOST targets, the above SNR cut has led to the exclusion 
of a significant fraction of stars. 
This may introduce some 
bias in favor of binaries since they tend to be brighter compared to their counterpart of similar spectral type.
Compared to the SDSS sample, the LAMOST sample contains more metal-rich stars from the Galactic disks.

To estimate the random errors of \feh~yielded by the LASP pipeline,
11,685 duplicate observations of comparable spectral SNRs of stars that fall in the
parameter ranges of the current sample
(4,300 $\le$ \teff~$\le$ 7,000\,K, \logg~$\ge$ 3.5\,dex, $-1.0$ $\le$ \feh~$\le$ 0.5\,dex)
are selected from the LAMOST DR1.
The random errors of \feh~ are fitted as a function of the SNR and \feh~using Eq.\,(4).
Only stars of SNRs lower than 50 are used in the fitting.
The resultant fit coefficients $a_0$ -- $a_5$ are 0.14, $-$0.063, 0.028,  $-$0.0039, 0.0013, and $4.4\times 10^{-5}$, respectively.
When assigning the random errors of \feh~for stars of SNRs higher than 50, the values given by the above fit for a SNR of 50 are used.
At SNR = 10, the random errors of \feh~are about 0.10 and 0.18\,dex for \feh~= 0 and $-$1\,dex, respectively.
At SNR = 50, the random errors are about 0.05 and 0.08\,dex for \feh~= 0 and $-$1\,dex, respectively.
For systematic errors, we have assumed,
\begin{equation} 
  \sigma_{\rm {sys}}({\rm [Fe/H]}) = 0.05 - 0.05 \times {\rm [Fe/H]}. 
\end{equation}
The values are assigned to be 0.05\,dex when \feh~$\ge$ 0.0\,dex.

\begin{figure}
\includegraphics[width=90mm]{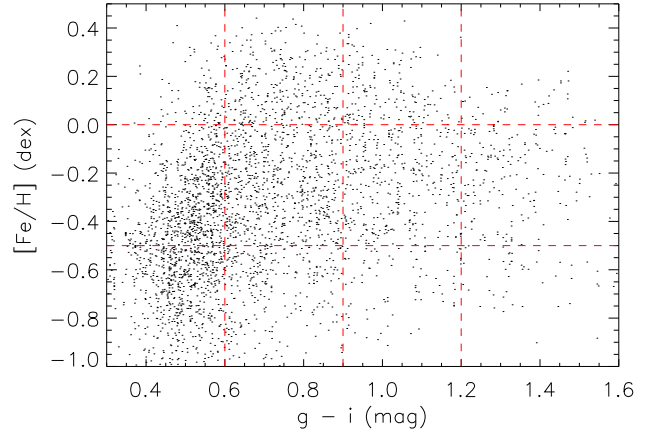}
\caption{
Distribution of the selected LAMOST DR1 stellar spectroscopic sample of Stripe 82 
in the $g - i$ and [Fe/H] plane.
}
\label{}
\end{figure}

\begin{figure}
\includegraphics[width=90mm]{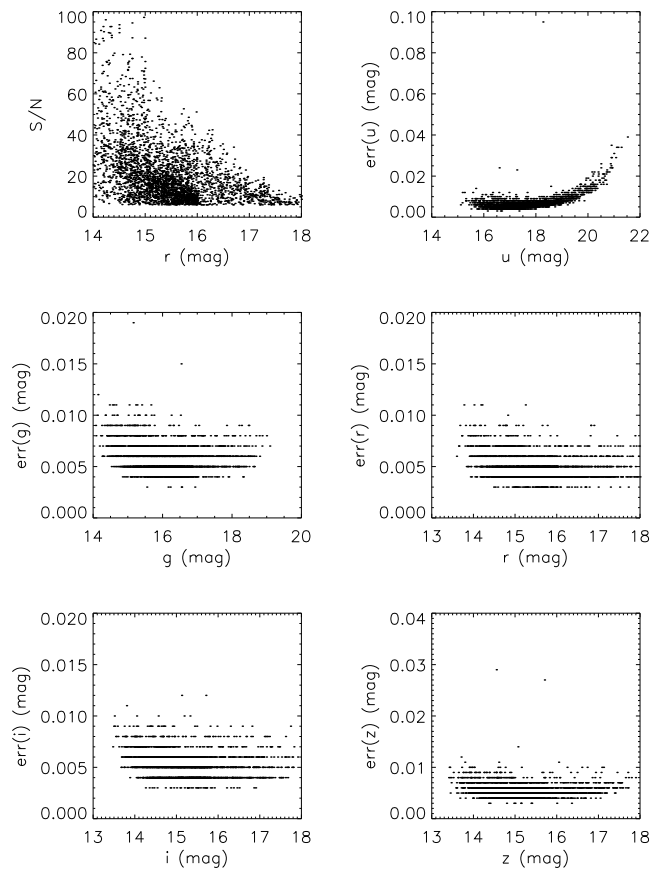}
\caption{
The spectral SNRs plotted against the $r$-band magnitudes
for the selected LAMOST DR1 stellar spectroscopic sample of Stripe 82 (top left panel).
The remaining panels show the photometric errors of the individual bands plotted against the magnitudes of the corresponding band.
}
\label{}
\end{figure}

\subsubsection{Results}
To account possible systematic differences in \feh~delivered by the SDSS DR9 and LAMOST DR1, 
we have derived the initial set of metallicity-dependent stellar loci separately 
for the LAMOST sample, using the same algorithm 
for the SDSS sample. The resultant fit coefficients are listed in the upper part of Table\,3.
Fig.\,16 compares the initial sets of stellar loci obtained for the SDSS and LAMOST samples.
The maximum differences are about 0.1\,mag in $u-g$ and about 0.01\,mag in other colors.

Similarly, two sets of MC simulations are performed for the LAMOST sample.
Each set of simulations is carried out ten times and then combined.
When determining the binary fraction,
the LAMOST sample has been divided into three bins of metallicity: $-1.0 \le$ \feh~$< -0.5$,
$-0.5 \le$ \feh~$< 0.0$, and $0.0 \le$ \feh~$< 0.5$\,dex.
For each bin of metallicity, the sample is further divided into three bins of color $g-i$:
$0.3 \le g-i <0.6$, $0.6 \le g-i < 0.9$, and $0.9 \le g-i \le 1.6$\,mag. 

The final set of stellar loci for the LAMOST sample are plotted in Fig.\,16.
Compared to the initial set, the differences in colors $u-g$, $g-r$, $r-i$, and $i-z$ are
respectively within the ranges of $-0.1$  -- 31.4, $-1.7$ -- 3.1, $-3.1$ -- 1.7, 
and $-3.2$ -- $-0.1$\,mmag for \feh~=~$0.0$\,dex, and
respectively within the ranges of $-10.6$  -- 49.3, $-1.7$ -- 2.1, $-2.1$ -- 1.7, 
and $-3.8$ -- $-1.9$\,mmag for \feh~=~$-1.0$\,dex.

Figs.\,17, 18, and 19 show 
the distributions of the observed residuals and of the best fit models for the LAMOST whole- and sub-samples
of individual bins of $g - i$ color and metallicity, for colors $u-g$, $g-r$, and $r-i$, respectively.
The ranges of color and metallicity of the bin, the resultant binary fraction and error, and the associated value
of minimum $\chi^2$ are marked. The fits are reasonably good. The typical minimum $\chi^2$ values are between 1.2 -- 3.0,
0.7 -- 2.0 and 0.4 -- 5.0 in colors $u-g$, $g-r$ and $r-i$, respectively. Similar to the SDSS sample,
the fits in $g-r$ color are the best. 
The binary fractions derived from the data of individual colors, are listed in Table\,4 and plotted in Fig.\,20
as a function of color for the individual bins of metallicity.
The weighted means are also listed and plotted, using the same weighting method for the SDSS sample.

The results from the individual colors are consistent.
The binary fractions of the LAMOST sample inferred from colors $u-g$, $g-r$, and $r-i$ are
$36\%\pm9\%$, $36\%\pm7\%$, and $46\%\pm7\%$,  respectively, yielding a weighted mean of $39\%\pm4\%$.
Similar to the SDSS sample, the binary fraction of field FGK dwarfs is also found to
decrease towards redder colors. The corresponding fractions are $57\%\pm12\%$, $42\%\pm6\%$, and $24\%\pm5\%$,
for stars of $g-i$ colors between 0.3 -- 0.6, 0.6 -- 0.9, and 0.9 -- 1.6\,mag, respectively.
Similarly, the binary fraction for field FGK dwarfs is found to decrease with increasing metallicity.
For stars of metallicities between  0.0 -- 0.5, $-0.5$ -- 0.0, and $-1.0$ -- $-0.5$\,dex,
the inferred binary fractions are $29\%\pm7\%$, $36\%\pm5\%$, and $48\%\pm11\%$, respectively

Fig.\,21 compares the binary fractions deduced from the SDSS and LAMOST samples.
The results are consistent within the uncertainties.
Both samples show the same trend:
the fraction of binaries is smaller for stars of redder colors and higher metallicities.

\begin{table}
\centering
\caption{Fit coefficients of the stellar color loci for the LAMOST sample.}
\label{}
\begin{tabular}{lrrrr} \hline\hline
Coeff. & $u-g^{a)}$ & $g-r^{b)}$ & $r-i^{b)}$ & $i-z^{b)}$  \\\hline
\noalign{\vskip2pt} \multicolumn{5}{c}{Initial set}  \\
$a_{0}$ &   1.2509 &   0.0399 &  $-$0.0399 &  $-$0.0698 \\
$a_{1}$ &  $-$0.4639 &  $-$0.0137 &   0.0137 &   0.0038 \\
$a_{2}$ &  $-$0.3676 &   0.0054 &  $-$0.0054 &  $-$0.0200 \\
$a_{3}$ &  $-$0.2515 &   0.0044 &  $-$0.0044 &  $-$0.0164 \\
$a_{4}$ &  $-$0.0599 &   0.6546 &   0.3454 &   0.1635 \\
$a_{5}$ &  $-$2.1854 &   0.0858 &  $-$0.0858 &  $-$0.0364 \\
$a_{6}$ &   2.1414 &   0.0072 &  $-$0.0072 &  $-$0.0048 \\
$a_{7}$ &   0.8310 &   0.0991 &  $-$0.0991 &   0.0565 \\
$a_{8}$ &   0.2444 &  $-$0.0370 &   0.0370 &   0.0013 \\
$a_{9}$ &   6.2080 &  $-$0.0411 &   0.0411 &  $-$0.0303 \\
$a_{10}$ &  $-$1.8420 &    &    &    \\
$a_{11}$ &  $-$0.3554 &    &    &    \\
$a_{12}$ &  $-$4.1650 &    &    &    \\
$a_{13}$ &   0.4834 &    &    &    \\
$a_{14}$ &   0.9218 &    &    &    \\
\noalign{\vskip2pt} \multicolumn{5}{c}{Final set}  \\
$a_{0}$ &   1.2979 &   0.0314 &  $-$0.0314 &  $-$0.0675 \\
$a_{1}$ &  $-$0.4330 &  $-$0.0205 &   0.0205 &   0.0116 \\
$a_{2}$ &  $-$0.3489 &   0.0022 &  $-$0.0022 &  $-$0.0144 \\
$a_{3}$ &  $-$0.2437 &   0.0064 &  $-$0.0064 &  $-$0.0150 \\
$a_{4}$ &  $-$0.0595 &   0.6832 &   0.3168 &   0.1548 \\
$a_{5}$ &  $-$2.5446 &   0.0993 &  $-$0.0993 &  $-$0.0515 \\
$a_{6}$ &   2.0669 &   0.0140 &  $-$0.0140 &  $-$0.0126 \\
$a_{7}$ &   0.7893 &   0.0775 &  $-$0.0775 &   0.0577 \\
$a_{8}$ &   0.2447 &  $-$0.0409 &   0.0409 &   0.0055 \\
$a_{9}$ &   7.0989 &  $-$0.0364 &   0.0364 &  $-$0.0284 \\
$a_{10}$ &  $-$1.7862 &    &    &    \\
$a_{11}$ &  $-$0.3094 &    &    &    \\
$a_{12}$ &  $-$4.9581 &    &    &    \\
$a_{13}$ &   0.4808 &    &    &    \\
$a_{14}$ &   1.1548 &    &    &    \\
\hline
\end{tabular}
\begin{description}
\item[$^{a)}$]  $f(x,y)=a_0+a_1*y+a_2*y^2+a_3*y^3+a_4*y^4+a_5*x+$ \\
             $ a_6*x*y+a_7*x*y^2+a_8*x*y^3+a_9*x^2+a_{10}*y*x^2+$ \\
             $a_{11}*y^2*x^2+a_{12}*x^3+a_{13}*y*x^3+a_{14}*x^4$, where $x$ denotes color $g-i$ and $y$ denotes metallicity \feh.
\item[$^{b)}$]  $f(x,y)=a_0+a_1*y+a_2*y^2+a_3*y^3+a_4*x+a_5*x*y+a_6*x*y^2+ $ \\
             $a_7*x^2+a_8*y*x^2+a_9*x^3$ , where $x$ denotes color $g-i$ and $y$ denotes metallicity \feh.
\end{description}
\end{table}

\begin{figure}
\includegraphics[width=90mm]{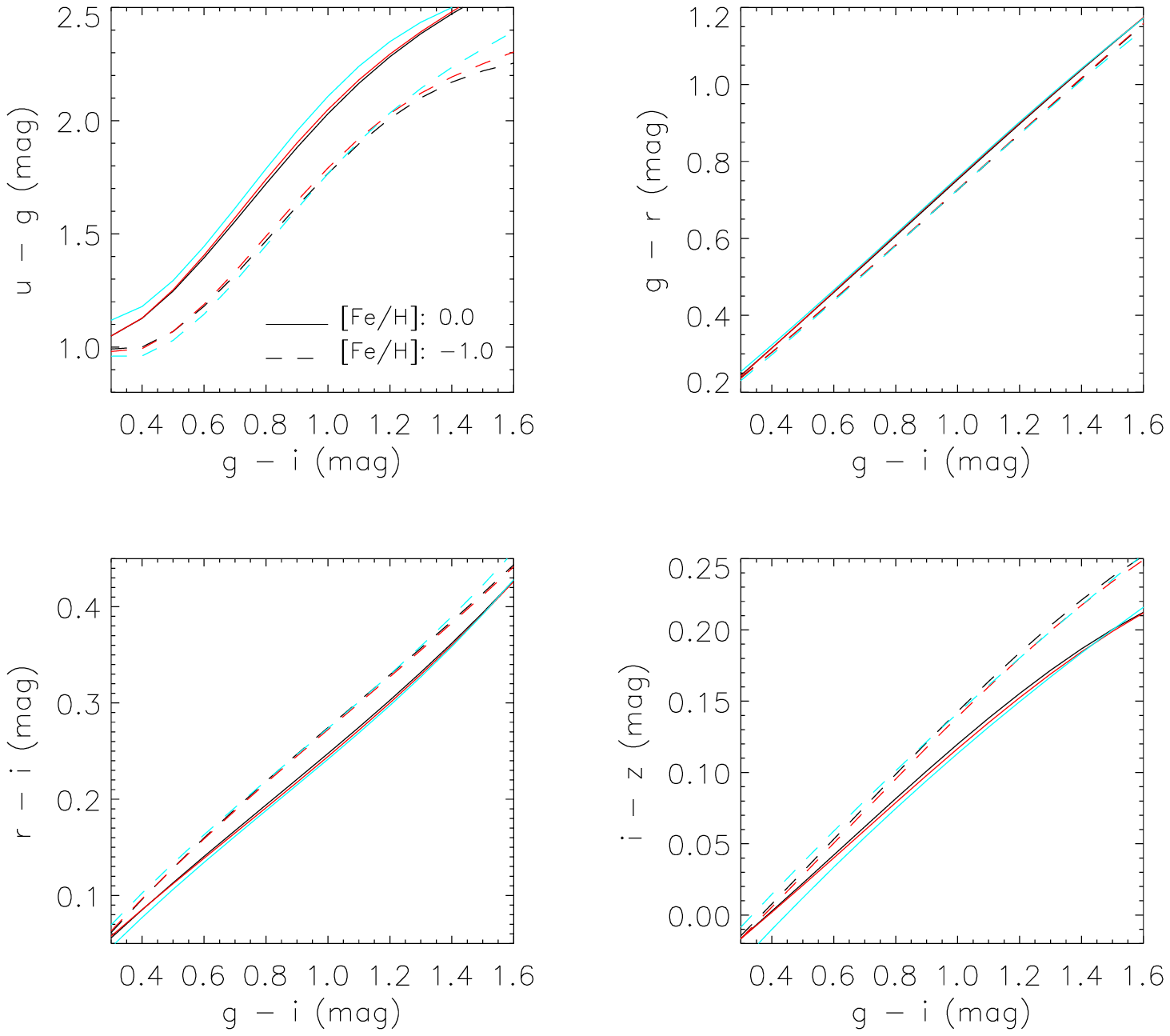}
\caption{
Initial (black) and final (red) sets of stellar color loci of 
\feh~=~0.0\,dex (solid lines) and $-$1.0\,dex (dashed lines),
obtained from the LAMOST sample. 
For comparison, the initial sets of loci deduced from the SDSS sample (cyan) are 
also overplotted.
}
\label{}
\end{figure}

\begin{figure*}\centering
\includegraphics[width=150mm]{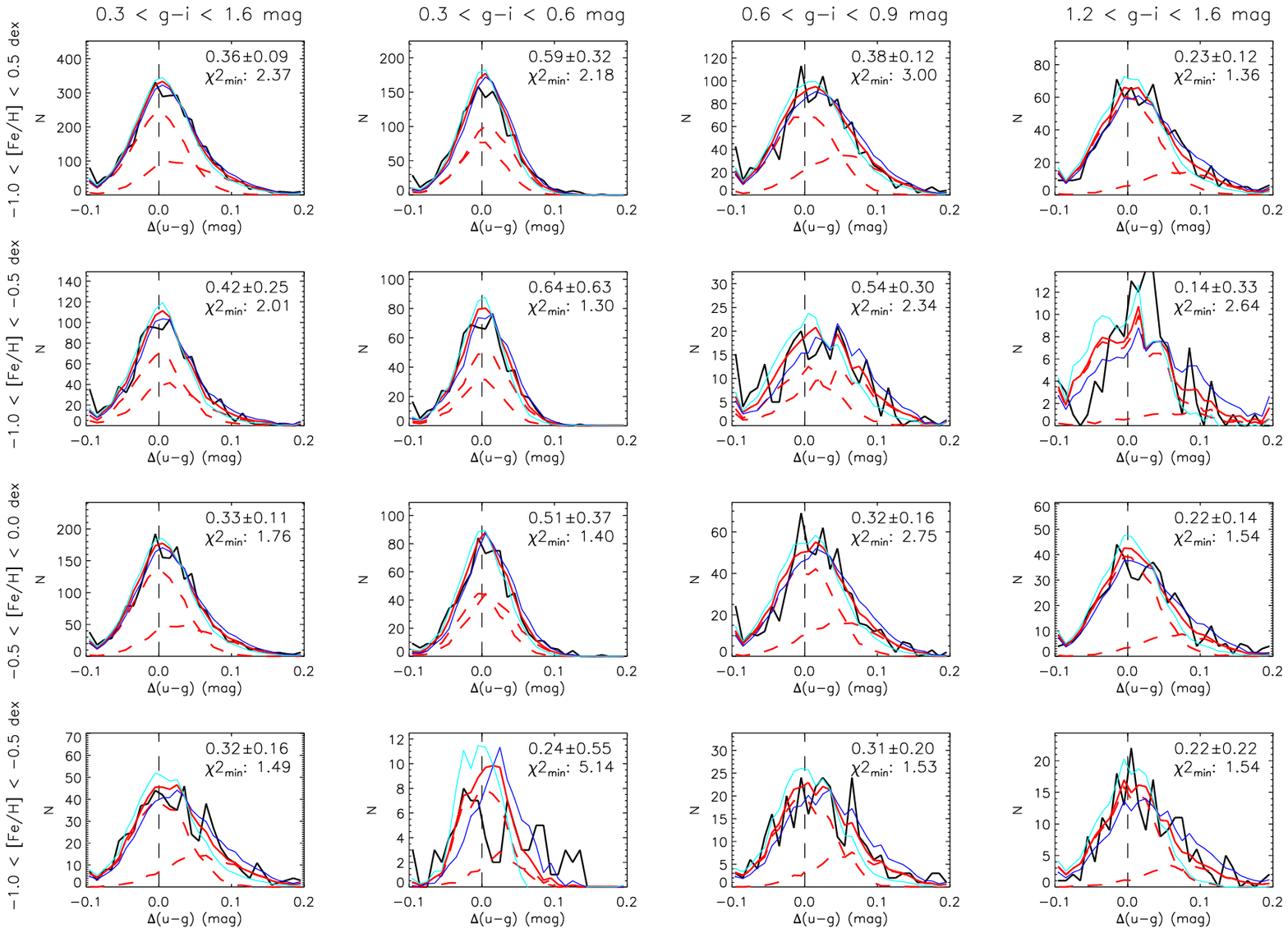}
\caption{
Same as Fig.\,10 but for the LAMOST sample and for residuals in $u-g$ color.
}
\label{}
\end{figure*}

\begin{figure*}\centering
\includegraphics[width=150mm]{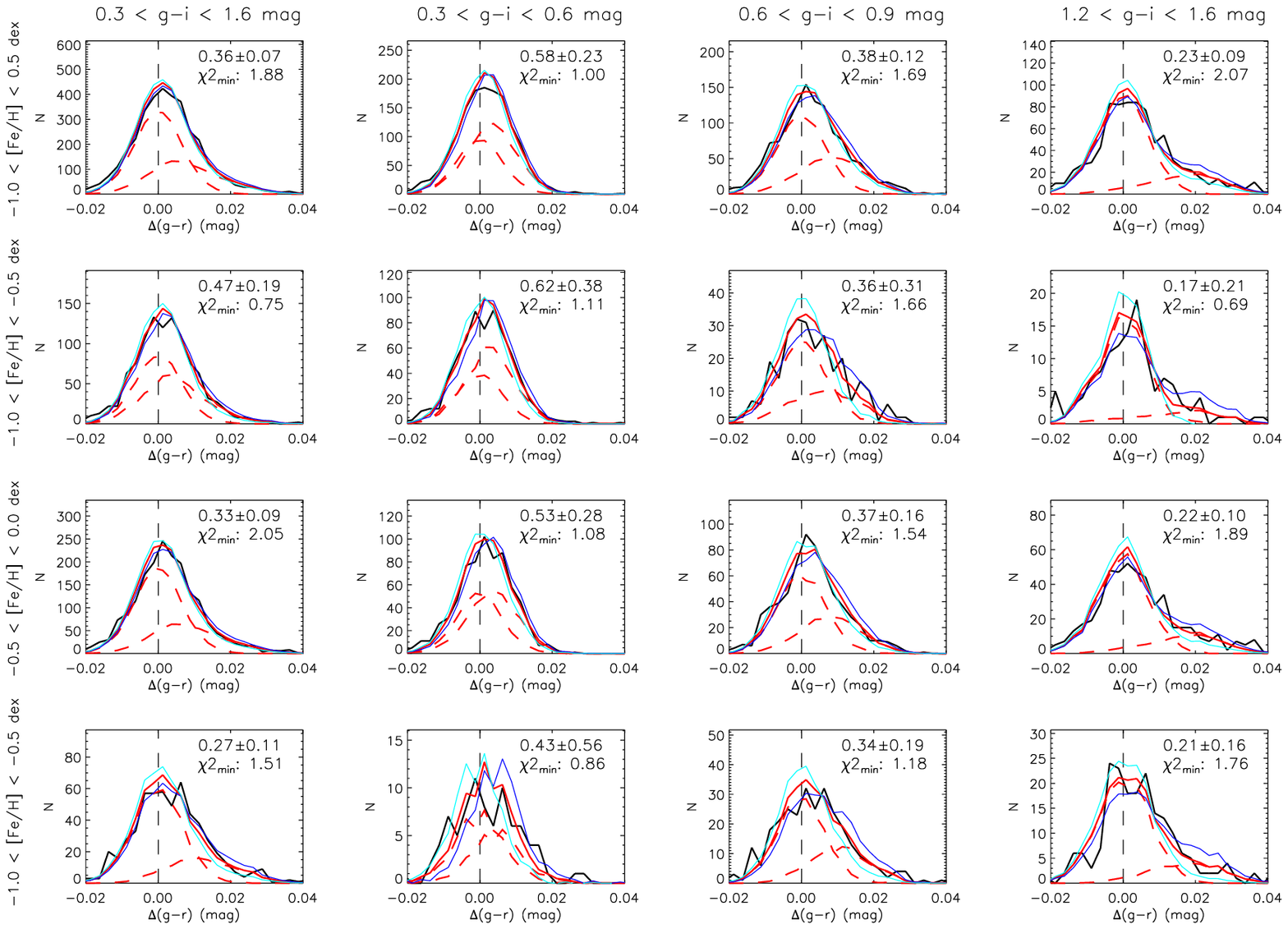}
\caption{
Same as Fig.\,10 but for the LAMOST sample and for residuals in $g-r$ color.
}
\label{}
\end{figure*}

\begin{figure*}\centering
\includegraphics[width=150mm]{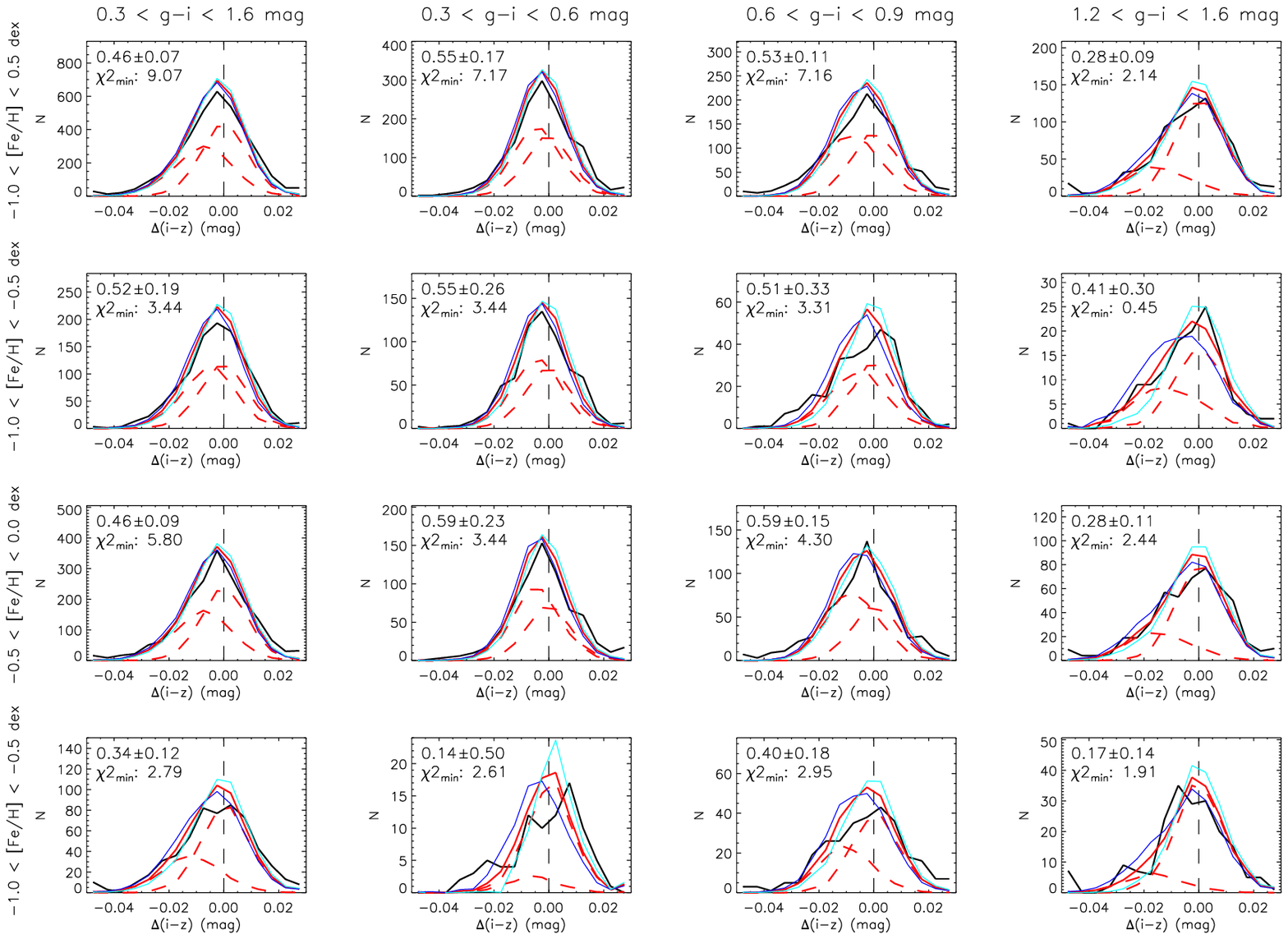}
\caption{
Same as Fig.\,10 but for the LAMOST sample and for residuals in $i-z$ color.
}
\label{}
\end{figure*}

\begin{figure*}\centering
\includegraphics[width=150mm]{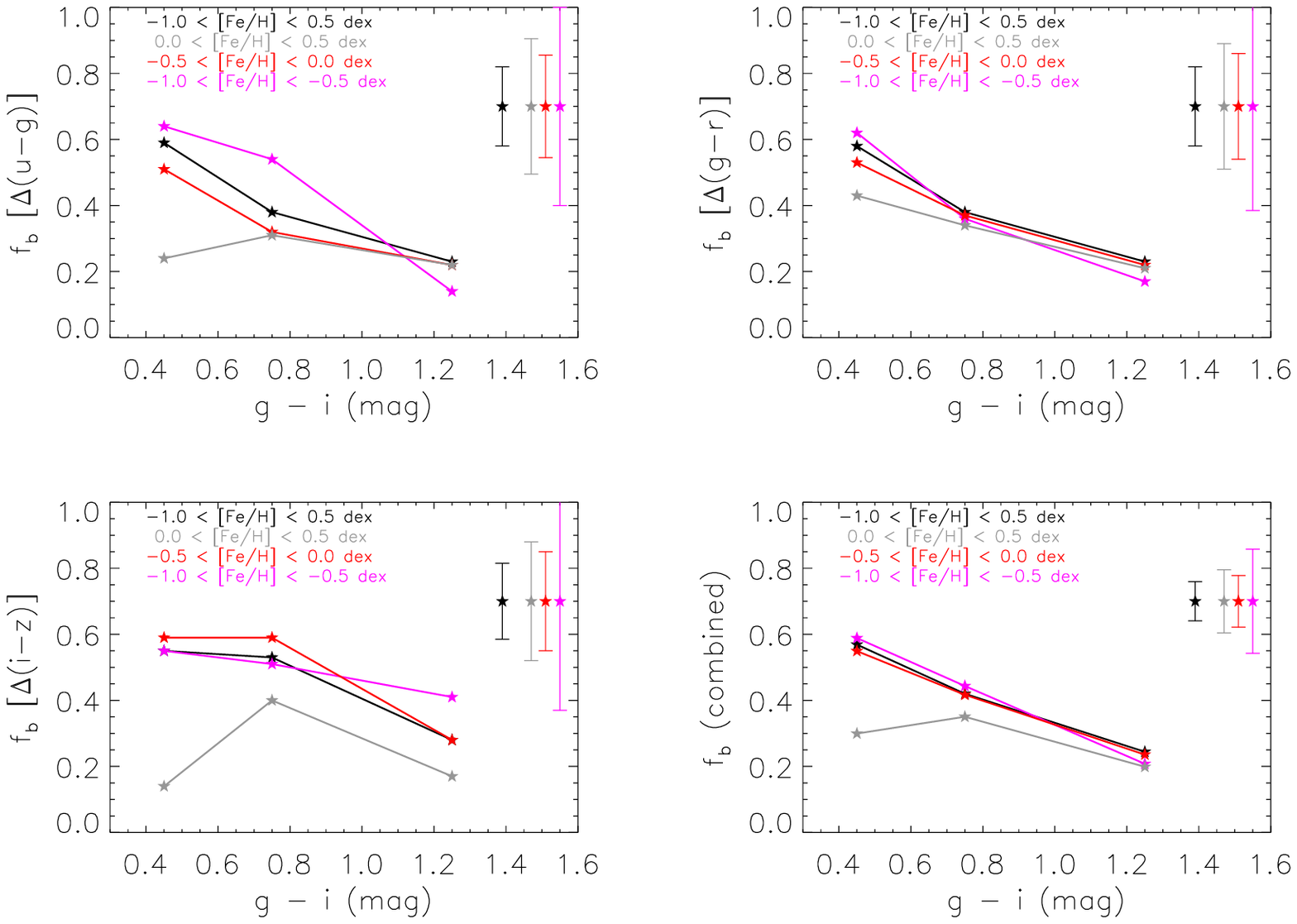}
\caption{
Binary fractions derived from the residuals in colors $u-g$ (top left), $g-r$ (top right),
$i-z$ (bottom left) and from the combined data of all three colors (bottom right) for field FGK stars of the LAMOST sample,
plotted against $g-i$ color for the individual bins of color and metallicity.
The typical error-bars are marked in the top-right corner of each panel.
}
\label{}
\end{figure*}

\begin{table*} 
\centering
\caption{Binary fractions of field FGK stars of the LAMOST sample. }
\label{}
\begin{tabular}{lcccc} \hline\hline
 &  $0.3 \le g-i \le 1.6$ &  $0.3 \le g-i < 0.6$ & $0.6 \le g-i < 0.9$ & $0.9 \le g-i \le 1.6$  \\
 & (mag) &  (mag) & (mag) & (mag)  \\
\noalign{\smallskip}\hline  \noalign{\vskip2pt} \multicolumn{5}{c}{Based on color $u-g$ }   \\
$-1.00 \le \feh < -0.50$\,dex & $0.42\pm0.26$ & $0.64\pm0.63$ & $0.55\pm0.30$ & $0.13\pm0.34$ \\
$-0.50 \le \feh <  0.00$\,dex & $0.34\pm0.11$ & $0.52\pm0.37$ & $0.33\pm0.16$ & $0.22\pm0.14$ \\
$-0.00 \le \feh \le 0.50$\,dex & $0.32\pm0.16$ & $0.25\pm0.56$ & $0.31\pm0.21$ & $0.22\pm0.22$ \\
$ -1.00 \le \feh \le 0.50$\,dex & $0.36\pm0.09$ & $0.59\pm0.32$ & $0.38\pm0.12$ & $0.23\pm0.12$ \\
\noalign{\smallskip}\hline  \noalign{\vskip2pt} \multicolumn{5}{c}{Based on color $g-r$  }   \\
$-1.00 \le \feh < -0.50$\,dex & $0.47\pm0.19$ & $0.62\pm0.38$ & $0.36\pm0.31$ & $0.17\pm0.21$ \\
$-0.50 \le \feh <  0.00$\,dex & $0.33\pm0.09$ & $0.53\pm0.28$ & $0.37\pm0.16$ & $0.22\pm0.11$ \\
$-0.00 \le \feh \le 0.50$\,dex & $0.27\pm0.11$ & $0.43\pm0.56$ & $0.34\pm0.19$ & $0.21\pm0.16$ \\
$ -1.00 \le \feh \le 0.50$\,dex & $0.36\pm0.07$ & $0.58\pm0.23$ & $0.38\pm0.12$ & $0.23\pm0.09$ \\
\noalign{\smallskip}\hline  \noalign{\vskip2pt} \multicolumn{5}{c}{Based on color $i-z$ }   \\
$-1.00 \le \feh < -0.50$\,dex & $0.52\pm0.19$ & $0.55\pm0.27$ & $0.51\pm0.33$ & $0.41\pm0.31$ \\
$-0.50 \le \feh <  0.00$\,dex & $0.46\pm0.09$ & $0.59\pm0.23$ & $0.59\pm0.15$ & $0.28\pm0.11$ \\
$-0.00 \le \feh \le 0.50$\,dex & $0.34\pm0.13$ & $0.15\pm0.49$ & $0.40\pm0.19$ & $0.17\pm0.14$ \\
$ -1.00 \le \feh \le 0.50$\,dex & $0.46\pm0.07$ & $0.55\pm0.17$ & $0.53\pm0.11$ & $0.28\pm0.09$ \\
\noalign{\smallskip}\hline  \noalign{\vskip2pt} \multicolumn{5}{c}{Combined }   \\
$-1.00 \le \feh < -0.50$\,dex & $0.48\pm0.11$ & $0.59\pm0.22$ & $0.45\pm0.16$ & $0.21\pm0.14$ \\
$-0.50 \le \feh <  0.00$\,dex & $0.36\pm0.05$ & $0.55\pm0.15$ & $0.42\pm0.08$ & $0.24\pm0.06$ \\
$-0.00 \le \feh \le 0.50$\,dex & $0.29\pm0.07$ & $0.30\pm0.27$ & $0.35\pm0.10$ & $0.20\pm0.09$ \\
$ -1.00 \le \feh \le 0.50$\,dex & $0.39\pm0.04$ & $0.57\pm0.12$ & $0.42\pm0.06$ & $0.24\pm0.05$ \\
\hline
\end{tabular}
\end{table*}

\begin{figure}
\includegraphics[width=90mm]{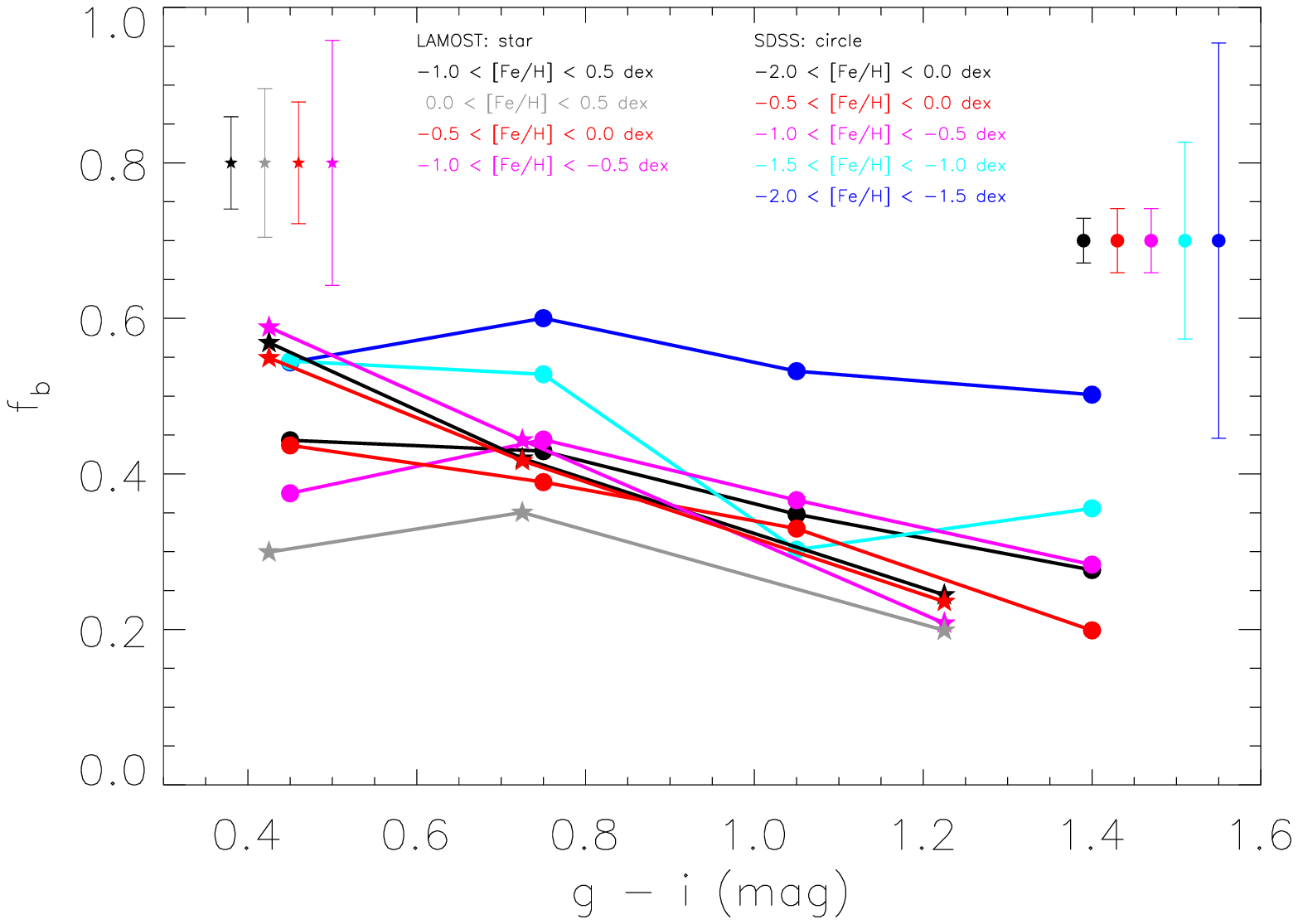}
\caption{
Binary fractions deduced from the SDSS (filled circles) and from the LAMOST (filled stars) 
samples plotted against $g-i$ color for the individual bins of color and metallicity.
The typical error-bars are marked.
}
\label{}
\end{figure}

\section{Binary fractions in Galactic disks and halo}
Earlier studies of the binary fraction are often  
limited to samples of the solar neighborhood, consisting mostly 
mental-rich stars from the Galactic thin disk. It is thus difficult to use the results to explore 
the binary fractions for different stellar populations, which 
are very important for understanding the formation and evolution of binaries in different environments.
The samples analyzed with the SLOT method in the current work probe a much larger and deeper volume
than previous studies, from a few hundred pc to over ten kpc, thus provide us a good opportunity 
to explore the fractions of binary stars in different stellar populations.

We have divided the SDSS sample stars into four populations based on their locations in the 
\feh~-- [$\alpha$/Fe] plane (Lee et al. 2011): thin disk stars (\feh $>$ $-$0.6\,dex, [$\alpha$/Fe] $\le$ 0.2\,dex),
thick disk stars (\feh~$>$ $-$1.0\,dex, [$\alpha$/Fe] $>$ 0.3\,dex), 
stars in the transition zone between the thin and thick disks (\feh~$>$ $-$0.6\,dex, 0.2 $<$ [$\alpha$/Fe] $\le$ 0.3\,dex), 
and halo stars (\feh~$\le$ $-$1.0\,dex). 
Like \feh~, here the [$\alpha$/Fe] are those yielded by the SSPP pipeline.
To ensure robust estimates of [$\alpha$/Fe], an additional spectral SNR cut of higher than 20 is imposed for the selections
of stars except for halo stars.
Since the binary fraction has been found to vary with color, 
to minimize the color effects, only stars of $g-i$ color bluer than 0.9\,mag are included.
The resultant distributions of stars thus selected for the four populations 
in the $g-i$ -- \feh~ and \feh~-- [$\alpha$/Fe] planes are shown in Fig.\,22.

Using the same simulated SDSS samples described in Section\,3.1, the binary fractions for the four populations are determined.
Figs.\,23, 24, and 25 show the distributions of the observed residuals and of the best fit models for 
the four populations in individual bins of $g - i$ color, for colors $u-g$, $g-r$, and $r-i$, respectively. 
The resultant binary fraction and error, and the associated value of minimum $\chi^2$ are marked for each panel.
The typical minimum $\chi^2$ values are between 1.0 -- 3.0,
0.5 -- 2.0 and 1.0 -- 3.0 for colors $u-g$, $g-r$ and $r-i$, respectively. 
The binary fractions, yielded by stars of different colors, are listed in Table\,5.
The weighted means are also listed, with the same weighting algorithm adopted in \S{3.1.2}.

The binary fractions deduced for stars of populations of the Galactic thin and thick disks, of 
the transition zone between them, and of the halo 
are $39\%\pm6\%$, $39\%\pm5\%$, $35\%\pm4\%$, and $55\%\pm10\%$, respectively.
The results suggest that the Galactic thin and thick disks have comparable binary fractions, 
whereas the Galactic halo contains a significantly larger binary fraction. 
The trend is the same for results obtained from the individual colors and from the individual color bin.
The results indicate that halo stars are formed in environment different to that of disk stars.

\begin{table} 
\centering
\caption{Binary fractions of field FGK stars in Galactic disks and halo.}
\label{}
\begin{tabular}{lccc} \hline\hline
 &  $0.3 \le g-i \le 0.9$ &  $0.3 \le g-i < 0.6$ & $0.6 \le g-i < 0.9$ \\
 &  (mag) & (mag) & (mag) \\
\noalign{\smallskip}\hline  \noalign{\vskip2pt} \multicolumn{4}{c}{Based on color $u-g$}   \\
Thin disk$^a$ & $0.36\pm0.10$ & $0.33\pm0.18$ & $0.46\pm0.12$ \\
Tran. disk$^b$ & $0.39\pm0.10$ & $0.34\pm0.18$ & $0.45\pm0.11$ \\
Thick disk$^c$  & $0.24\pm0.07$ & $0.29\pm0.15$ & $0.25\pm0.08$ \\
Halo & $0.52\pm0.17$ & $0.54\pm0.18$ & $0.57\pm0.22$ \\
\noalign{\smallskip}\hline  \noalign{\vskip2pt} \multicolumn{4}{c}{Based on color $g-r$}   \\
Thin disk$^a$ & $0.38\pm0.12$ & $0.29\pm0.21$ & $0.44\pm0.12$ \\
Tran. disk$^b$ & $0.39\pm0.11$ & $0.34\pm0.19$ & $0.44\pm0.13$ \\
Thick disk$^c$  & $0.40\pm0.08$ & $0.34\pm0.14$ & $0.43\pm0.09$ \\
Halo & $0.58\pm0.23$ & $0.51\pm0.33$ & $0.55\pm0.22$ \\
\noalign{\smallskip}\hline  \noalign{\vskip2pt} \multicolumn{4}{c}{Based on color $i-z$ }   \\
Thin disk$^a$ & $0.42\pm0.10$ & $0.39\pm0.17$ & $0.45\pm0.12$ \\
Tran. disk$^b$ & $0.40\pm0.10$ & $0.40\pm0.18$ & $0.40\pm0.12$ \\
Thick disk$^c$  & $0.39\pm0.07$ & $0.36\pm0.13$ & $0.41\pm0.09$ \\
Halo & $0.54\pm0.19$ & $0.43\pm0.22$ & $0.59\pm0.29$ \\
\noalign{\smallskip}\hline  \noalign{\vskip2pt} \multicolumn{4}{c}{Combined }   \\
Thin disk$^a$ & $0.39\pm0.06$ & $0.33\pm0.10$ & $0.45\pm0.06$ \\
Tran. disk$^b$ & $0.39\pm0.05$ & $0.36\pm0.09$ & $0.43\pm0.06$ \\
Thick disk$^c$ & $0.35\pm0.04$ & $0.33\pm0.07$ & $0.37\pm0.04$ \\
Halo$^d$ & $0.55\pm0.10$ & $0.50\pm0.14$ & $0.56\pm0.12$ \\
\hline
\end{tabular}
\begin{description}
\item[$^a$] S/N $\ge$ 20, \feh~ $>$ $-$0.6\,dex, [$\alpha$/Fe] $\le$ 0.2\,dex. 
\item[$^b$] S/N $\ge$ 20, \feh~ $>$ $-$0.6\,dex, 0.2 $<$ [$\alpha$/Fe] $\le$ 0.3\,dex.
\item[$^c$] S/N $\ge$ 20, \feh~ $>$ $-$1.0\,dex, [$\alpha$/Fe] $>$ 0.3\,dex.
\item[$^d$] S/N $\ge$ 10, \feh~$\le$ $-$1.0\,dex.
\end{description} 
\end{table}

\begin{figure}
\includegraphics[width=90mm]{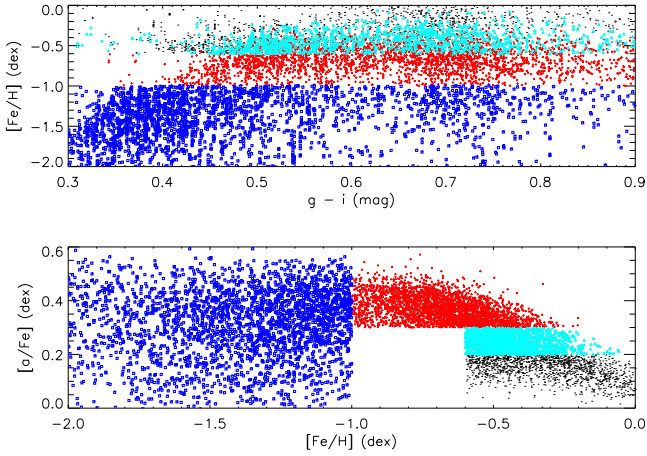}
\caption{
Distributions of stars of different stellar populations 
(black: thin disk; cyan: transition disk; red: thick disk; blue: halo) of
the SDSS sample in the \feh~-- $g-i$ (top panel) and [$\alpha$/Fe] -- \feh~(bottom panel) planes. 
}
\label{}
\end{figure}

\begin{figure*}\centering
\includegraphics[width=150mm]{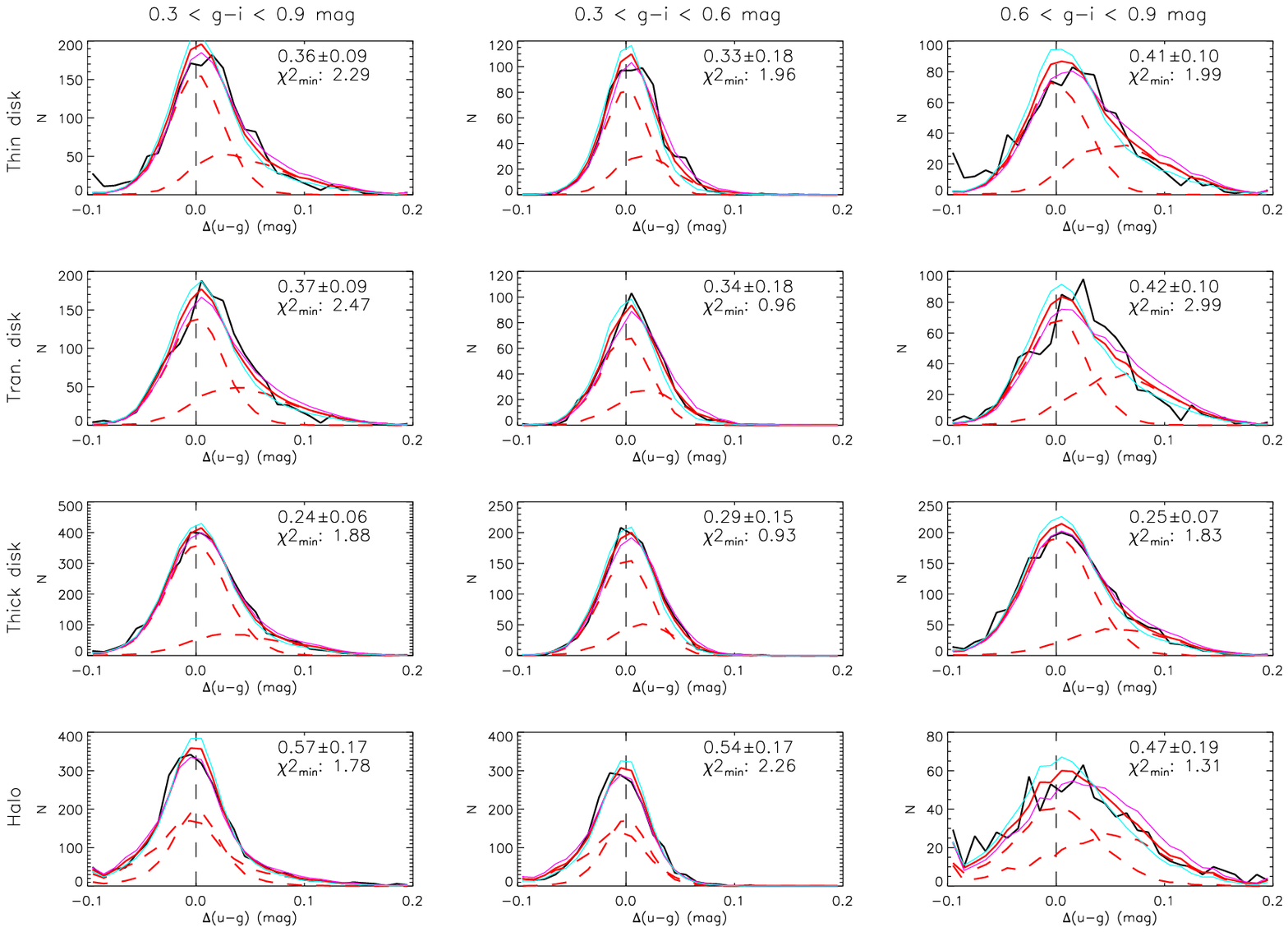}
\caption{
Same as Fig.\,10 for the SDSS sample of different stellar populations in $u-g$ color. 
}
\label{}
\end{figure*}

\begin{figure*}
\centering
\includegraphics[width=150mm]{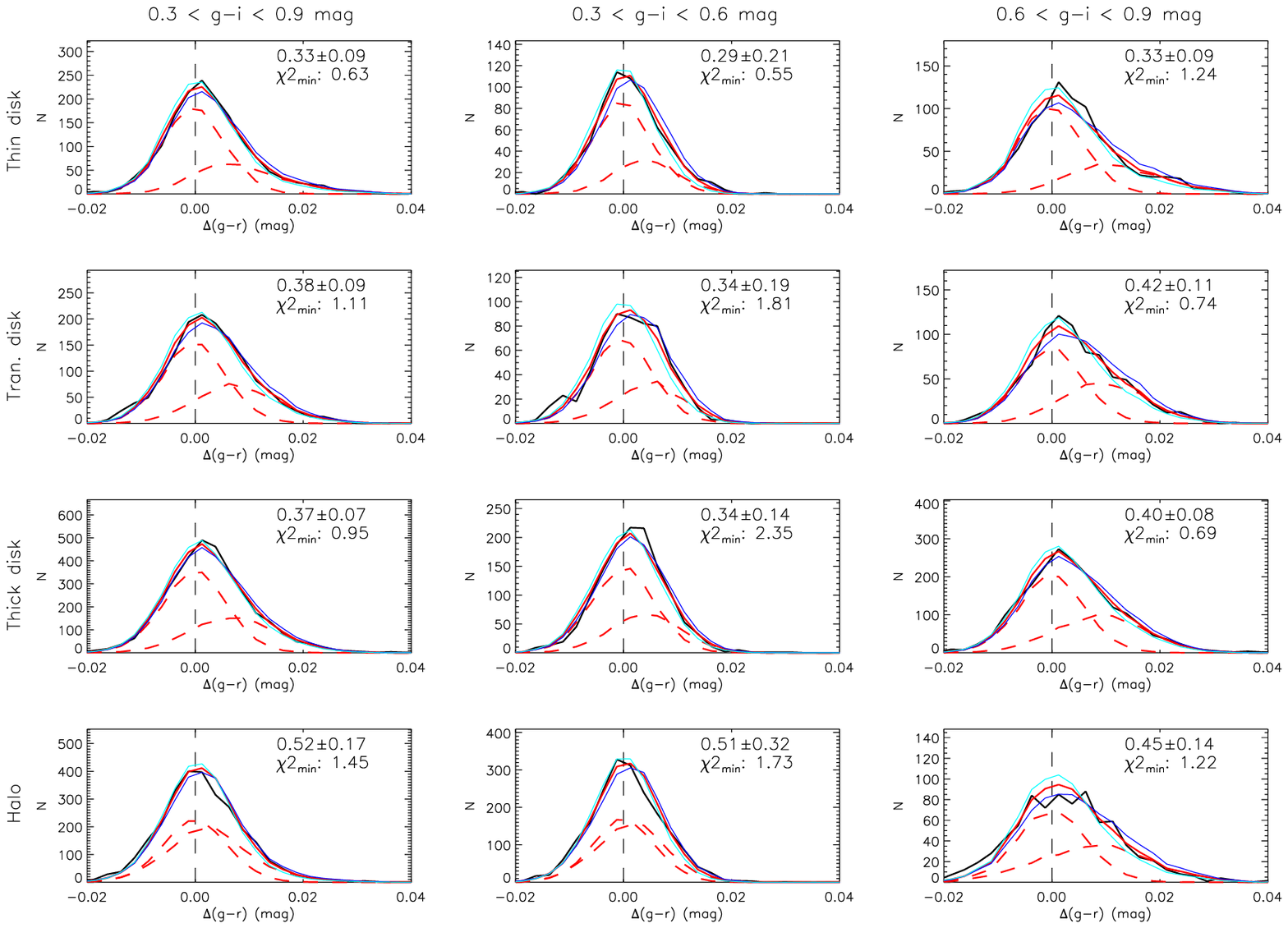}
\caption{
Same as Fig.\,10 for the SDSS sample of different stellar populations in $g-r$ color. 
}
\label{}
\end{figure*}

\begin{figure*}\centering
\includegraphics[width=150mm]{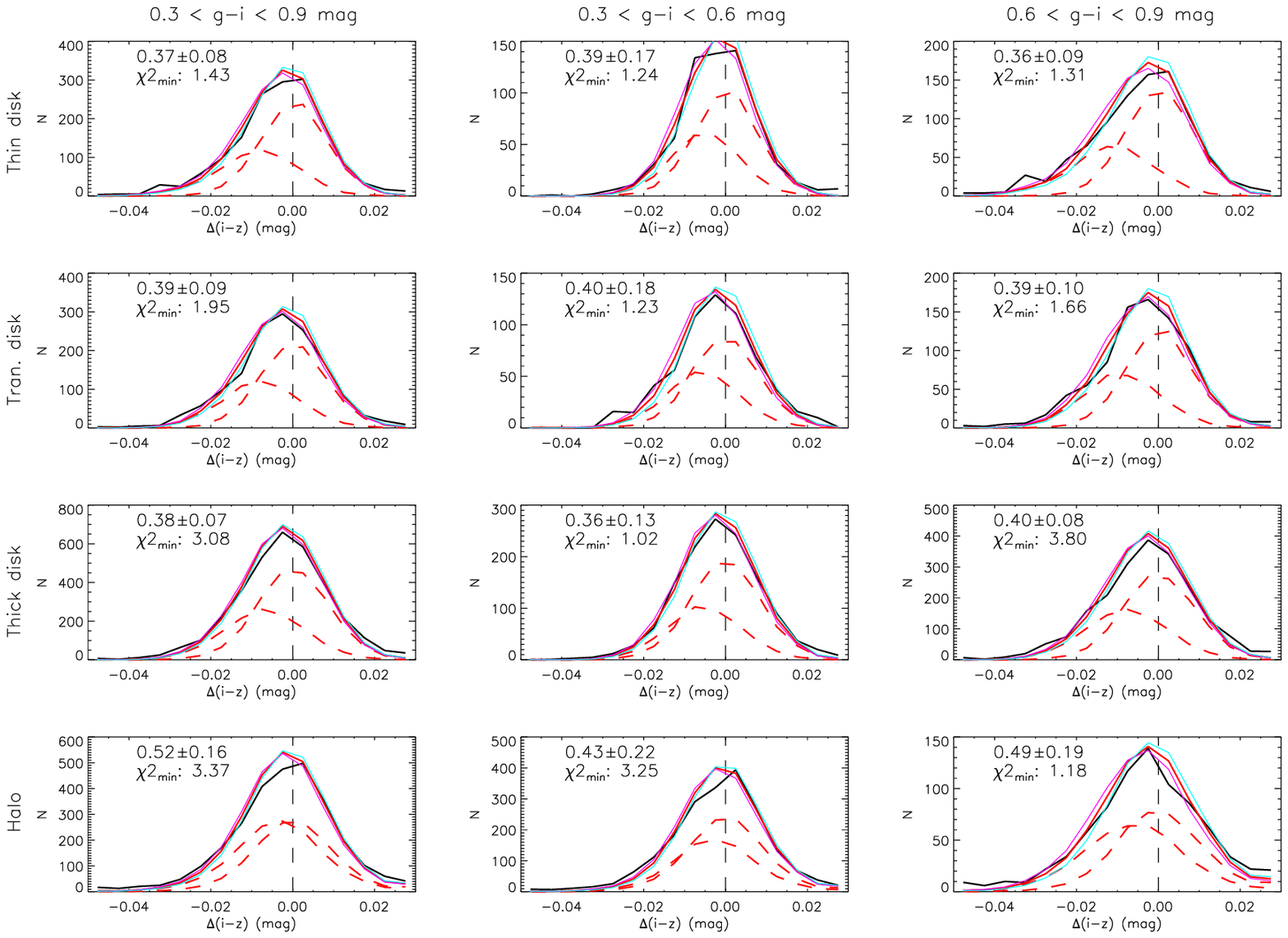}
\caption{
Same as Fig.\,10 for the SDSS sample of different stellar populations in $i-z$ color.
}
\label{}
\end{figure*}

\section{Conclusions and discussions}
In Paper\,I, by combining the spectroscopic information and recently re-calibrated photometry of
the SDSS Stripe 82, we have built a large, clean sample of main sequence stars with accurate colors
and  well determined metallicities, and demonstrated that 
the $u-g$, $g-r$, $r-i$ and  $i-z$ colors of MS stars are fully determined 
by their $g-i$ colors and metallicities. In other words, the intrinsic widths of the 
metallicity-dependent stellar color loci are essential zero.  
In this paper, we have shown that MS binaries do not follow the metallicity-dependent stellar loci 
defined by single stars. Binaries can be discriminated from single stars by their small deviations from the loci. 
The deviations from the metallicity-dependent stellar loci of an observed sample of MS stars are contributed by:
1) Uncertainties in the color measurements and calibrations; 2) Uncertainties in the metallicity determinations, and 3) 
The presence of binaries.
By modeling the observed deviations from the metallicity-dependent stellar loci,
we propose a SLOT method to determine the binary fraction of MS stars statistically.
The method is sensitive to neither the period nor
mass-ratio distributions of binaries, and applicable to large survey volumes.
With accurate colors measured by modern photometric surveys and robust estimates of metallicities
and surface gravities delivered by large scale spectroscopic surveys,
the SLOT method provides a promising tool to obtain model-free estimates of binary fraction 
for large numbers of stars of different populations.

We have applied the SLOT method to two samples of stars from the SDSS Stripe 82, 
constructed by combining the re-calibrated SDSS photometric data with respectively the 
spectroscopic information from the SDSS and LAMOST surveys.
The SDSS sample contains 14,650 MS stars of $0.3 \le g-i \le 1.6$\,mag and $-2.0 \le \feh \le 0.0$\,dex.
The LAMOST sample contains 3,827 MS stars of $0.3 \le g-i \le 1.6$\,mag and $-1.0 \le \feh \le 0.5$\,dex. 
The binary fractions deduced from residuals in colors $u-g$, $g-r$ and $i-z$ agree with each other.
For the whole SDSS sample, we find that the binary fraction of field FGK stars is 41\%$\pm$2\%.
The fraction is found to decrease towards redder stars (i.e. stars of later spectral types). 
The fractions are respectively $44\%\pm5\%$, $43\%\pm3\%$, $35\%\pm5\%$, and $28\%\pm6\%$,
for stars of $g-i$ colors between 0.3 -- 0.6, 0.6 -- 0.9, 0.9 -- 1.2, and 1.2 -- 1.6\,mag.
The fraction is also found to increase with decreasing metallicity.
The trend is likely stronger for stars of redder colors.
For stars of \feh~between $-0.5$ -- 0.0, $-1.0$ -- $-0.5$, $-1.5$ -- $-1.0$, and $-2.0$ -- $-1.5$\,dex,
the inferred binary fractions are $37\%\pm3\%$, $39\%\pm3\%$, $50\%\pm9\%$, and $53\%\pm20\%$, respectively.
We have further divided the SDSS sample into stars from the Galactic thin and thick disks, from the transition zone 
between them, and from the halo by their locations in the \feh~-- [$\alpha$/Fe] plane.
For the four populations, the binary fractions are $39\%\pm6\%$, $39\%\pm5\%$, $35\%\pm4\%$, and $55\%\pm10\%$, respectively,
suggesting that thin and thick disk stars have similar binary fractions, whereas for halo stars, the fraction seems to be significantly higher.
Applying the method to the LAMOST sample yields consistent results.
The binary fraction of the whole LAMOST sample is 39\%$\pm$4\%.
The fractions for stars of $g-i$ colors between 0.3 -- 0.6, 0.6 -- 0.9, and 0.9 -- 1.6\,mag 
are $57\%\pm12\%$, $42\%\pm6\%$, and $24\%\pm5\%$, respectively.
For stars of \feh~between 0.0 -- 0.5, $-0.5$ --  0.0, and $-1.0$ -- $-0.5$\,dex,
the inferred binary fractions are $29\%\pm7\%$, $36\%\pm5\%$, and $48\%\pm11\%$, respectively.

Given the inferred dependence of the binary fraction on stellar color and metallicity, 
the binary fraction inferred from samples of stars of a broad range of color and metallicity with the SLOT method,
for example, the whole SDSS and LAMOST samples in the current work,  
will depend on the color and metallicity distributions of the sample stars. 
However, for (sub-)samples of narrower color and metallicity ranges, for example, the SDSS and LAMOST subsamples of individual 
bins of color and metallicity in the current work, 
the derived binary fractions are barely affected by the color and metallicity distributions of the sample stars.
The binary fractions derived depend on, but are insensitive to, the assumed mass ratio distribution of binary stars. 
In this work, we have adopted a power law mass ratio distribution of index $\gamma = 0.3$. 
Tests with a completely different index of $\gamma = 1$ show that 
the inferred fractions decrease only by a few per cent, whereas the trends of variations of the fraction 
with stellar color and metallicity remain unchanged.
Although the SLOT method suffers from little bias, the samples could be biased.
The samples used in this work include only stars of spectral SNR higher than a specific value, 
thus may introduce a Malmquist bias that favors brighter binaries.
Given the good SNRs and the wide range of magnitude of the SDSS sample stars,
the effects of Malmquist bias for the SDSS sample are likely to be small.
For the LAMOST sample, the effects could be somewhat larger. 

In the current work, we have neglected binary stars composed of a normal star and a 
compact object, such as a white dwarf, a neutron stars, or a black hole.
However, the fraction of such systems should be negligible compared to binaries consisting of two normal stars.
In the current work, we have also neglected wide binaries that are spatially resolved in the SDSS images.
They would be missed by the method. However,  given that 
the median distances of the SDSS and LAMOST sample stars are respectively about 2.0 and 1.0 \,kpc and that 
90 per cent of the targets are at least 1.0 and 0.6\,kpc away,
the numbers of such systems that have been left out are likely to be very small (e.g., Sesar et al. 2008). 
Superpositions of two unassociated stars by chance are also neglected, considering the high Galactic latitudes of Stripe 82.
Systems consisting of triple or more stars are also neglected, which shall be regarded as 
binaries as the effects of the third and higher-order companions on the combined colors are negligible.  

The SLOT method proposed in this work is limited to stars spectroscopically targeted due to the requirement of 
robust estimates of metallicity and surface gravity.
Given the strong dependence of $u-g$ color on metallicity, 
it is worth exploring the dependence of colors $g-r$ and $i-z$ as a function of $g-i$ and $u-g$ in the future.
With sophisticated modeling, it is possible to extend the method to estimate binary fractions 
for a complete sample of stars using photometric data alone.  

The SLOT method is only applied to Stripe 82 in the current work, as Stripe 82 has extremely well-calibrated 
photometry with colors accurate to a few mmag. Applying the method to all stars spectroscopically targeted by the 
SDSS and LAMOST surveys, with photometry from the SDSS, is also possible, 
once the SDSS colors have been further re-calibrated using the stellar color regression method (Yuan et al. 2014a).

The SLOT method can be applied to star clusters using  accurate photometric data alone, 
as cluster member stars presumably have the same metallicity. By analyzing the 
offsets in colors as well as in brightness, one can constrain 
both the binary fraction and mass ratio distribution of cluster member stars.   
If accurate distances to individual field stars are available, for example from the Gaia satellite (Perryman et al. 2001),
the offsets in colors and  brightness can also be used to constrain
the binary fraction and mass ratio distribution for field stars.

Binary fraction for field stars can also be estimated by variations in stellar radial velocities 
from multi-epoch spectroscopy (Gao et al. 2014).
Thanks to the on-going LAMOST surveys, multi-epoch spectroscopic data have increased significantly. 
A significant fraction of stars targeted by the LAMOST has been observed twice or multiple times.  
By combining the offsets in colors and the variations in radial velocity, one can constrain both 
the binary fraction and orbital period distribution very well. 
In addition, a color offset -- velocity variation diagram will provide a 
simple classification scheme to discriminate  MS binaries, compact binaries with a luminous (hot) companion, and 
those with a faint companion, as shall be introduced in the third paper of this series (Yuan et al. 2014d; in preparation). 
By searching for MS stars that exhibit large variations in 
radial velocity but zero or unusual color offsets, one may have the possibility of finding a large number of 
stellar mass black hole- and neutron star binary systems in the Galaxy. 

\vspace{7mm} \noindent {\bf Acknowledgments}{
We would like to thank the referee for his/her helpful comments.
We would like to acknowledge the useful discussions with M.B.N. (Thijs) Kouwenhoven and Chengyuan Li.
This work is supported by National Key Basic Research Program of China 2014CB845700
and China Postdoctoral Science special Foundation 2014T70011.

Funding for SDSS-III has been provided by the Alfred P. Sloan Foundation, the
Participating Institutions, the National Science Foundation, and the U.S.
Department of Energy Office of Science. The SDSS-III web site is
http://www.sdss3.org/.

SDSS-III is managed by the Astrophysical Research Consortium for the
Participating Institutions of the SDSS-III Collaboration including the
University of Arizona, the Brazilian Participation Group, Brookhaven National
Laboratory, Carnegie Mellon University, University of Florida, the French
Participation Group, the German Participation Group, Harvard University, the
Instituto de Astrofisica de Canarias, the Michigan State/Notre Dame/JINA
Participation Group, Johns Hopkins University, Lawrence Berkeley National
Laboratory, Max Planck Institute for Astrophysics, Max Planck Institute for
Extraterrestrial Physics, New Mexico State University, New York University,
Ohio State University, Pennsylvania State University, University of Portsmouth,
Princeton University, the Spanish Participation Group, University of Tokyo,
University of Utah, Vanderbilt University, University of Virginia, University
of Washington, and Yale University.

Guoshoujing Telescope (the Large Sky Area Multi-Object Fiber Spectroscopic Telescope, LAMOST) is
a National Major Scientific Project built by the Chinese Academy of Sciences. Funding for the project
has been provided by the National Development and Reform Commission. LAMOST is operated and managed by the National
Astronomical Observatories, Chinese Academy of Sciences.
}

\label{lastpage}

\end{document}